\documentclass[12pt]{article}
\usepackage[reqno]{amsmath}
\usepackage{epsfig}
\usepackage{array}
\usepackage{float}
\usepackage{lscape,graphicx}
\usepackage{graphics}
\usepackage{amssymb}
\usepackage{color}

\def\ltap{\ \raisebox{-.4ex}{\rlap{$\sim$}} \raisebox{.4ex}{$<$}\ }
\def\gtap{\ \raisebox{-.4ex}{\rlap{$\sim$}} \raisebox{.4ex}{$>$}\ }

\newcommand{\betabeta}{\mbox{$(\beta \beta)_{0 \nu}  $}}

\newcommand{\meff}{\mbox{$\left|  \langle \!  m \!  \rangle \right| $}}

\newcommand{\eV}{\mbox{$~{\rm eV}$}}

\newcommand{\bea}{\begin{equation}\begin{array}{c}}
\newcommand{\eea}{\end{array}\end{equation}}
\newcommand{\ea}{\end{array}}

\newcommand{\beq}{\begin{equation}}
\newcommand{\eeq}{\end{equation}}
\newcommand{\bad}{\begin{array}{ccc}}
\newcommand{\dmsol}{\mbox{$\Delta m^2_{\odot}$}}
\newcommand{\dma}{\mbox{$\Delta m^2_{\rm A}$ }}
\newcommand{\mefff}{\mbox{$ \langle \! m \! \rangle $}}
\newcommand{\ba}{\begin{array}{c}}
\newcommand{\half}{\frac{1}{2}}
\newcommand{\diag}{{\rm diag}}

\begin{document}
\hfill{{\small Ref. FLAVOUR(267104)-ERC-15}}

\hfill{{\small Ref. SISSA 10/2012/EP}}

\hfill{{\small Ref. TUM-HEP 837/12}}

\hfill{{\small Ref. CFTP/12-007}}


\begin{center}
\mathversion{bold}
{\bf{\large The $\mu - e$ Conversion in Nuclei, $\mu \to e \gamma$,
$\mu \to 3e$ Decays and TeV Scale See-Saw Scenarios of
Neutrino Mass Generation}}
\mathversion{normal}

\vspace{0.4cm}
D. N. Dinh$\mbox{}^{a,b)}$,
A. Ibarra$\mbox{}^{c)}$,
E. Molinaro$\mbox{}^{d)}$ and
S. T. Petcov$\mbox{}^{a,e)}$~\footnote{Also at: Institute of Nuclear Research and
Nuclear Energy, Bulgarian Academy of Sciences, 1784 Sofia, Bulgaria}

\vspace{0.2cm}
$\mbox{}^{a)}${\em  SISSA and INFN-Sezione di Trieste, \\
Via Bonomea 265, 34136 Trieste, Italy.\\}

\vspace{0.1cm}
$\mbox{}^{b)}${\em Institute of Physics, 10 Dao Tan, Hanoi, Vietnam.\\
}

\vspace{0.1cm}
$\mbox{}^{c)}${\em Physik-Department T30d, Technische Universit\"at M\"unchen,\\ James-Franck-Stra{\ss}e, 85748 Garching, Germany.\\}

\vspace{0.1cm}
$\mbox{}^{d)}${\em  Centro de F\'{i}sica Te\'{o}rica de Part\'{i}culas, \\
Instituto Superior T\'{e}cnico, Technical University of Lisbon, \\ 1049-001,
Lisboa, Portugal. \\}

\vspace{0.1cm}
$\mbox{}^{e)}${\em Kavli IPMU, University of Tokyo (WPI), Tokyo, Japan.\\}

\end{center}

\begin{abstract}
We perform a detailed analysis of lepton flavour violation  (LFV) within minimal see-saw type extensions of the Standard Model (SM),
which give a viable mechanism of neutrino mass generation and provide new particle content at the  electroweak
scale. We focus, mainly, on predictions and constraints set on each scenario from
$\mu\to e \gamma$, $\mu\to 3e$ and $\mu-e$ conversion in the nuclei. In this class of models, the flavour structure of the
Yukawa couplings between  the additional scalar and fermion representations and the SM leptons is
highly constrained by neutrino oscillation measurements. In particular, we show that in some regions of the parameters
space of type I and type II  see-saw models, the Dirac and Majorana phases of the neutrino mixing matrix,
the ordering and hierarchy of the active neutrino mass spectrum as well as the value of the reactor mixing angle $\theta_{13}$
may considerably affect the size of the LFV observables.
The interplay of the latter clearly allows  to discriminate among the different low energy see-saw possibilities.

\end{abstract}

\section{Introduction}

After several decades of neutrino experiments, a clear
quantitative picture of the neutrino oscillation
parameters is gradually emerging (see, e.g. \cite{PDG10}).
The Super-Kamiokande collaboration established that the
atmospheric neutrino mass squared
splitting is $|\Delta m_A^2|\sim {\cal O}(10^{-3}\eV^2)$
and that the corresponding mixing angle is large,
possibly maximal $\theta_{23}\cong \pi/4$
~\cite{Fukuda:1998mi}.
The data from SNO, Super-Kamiokande and KamLAND
experiments \cite{ahmad01,fukuda01,KL162}
allowed to established the large mixing angle
solution as
a unique solution of the long standing
solar neutrino problem, with a solar
neutrino mass squared splitting
$\Delta m^{2}_{\odot} \sim {\cal O}( 10^{-5}\eV^2)$ and mixing angle
$\theta_{12} \cong \arcsin(\sqrt{0.3})$.
A series of subsequent
experiments, using reactor and accelerator neutrinos, have pinned
down the atmospheric and solar neutrino oscillation parameters with
a few to several percent accuracy,
as summarized in Table \ref{tab:tabdata-1106}.

  Furthermore, in June of 2011 the T2K collaboration  reported
\cite{Abe:2011sj} evidence at $2.5\sigma$
for a non-zero value of the angle $\theta_{13}$.
Subsequently the MINOS \cite{MINOS240611}
and Double Chooz \cite{DChooz2011}
collaborations also reported evidence
for $\theta_{13}\neq 0$,
although with a smaller statistical
significance. Global
analyses of the neutrino oscillation data, including
the data from the T2K and MINOS experiments, performed
in \cite{Fogli:2011qn,Schwetz11} showed  that actually
$\sin\theta_{13}\neq 0$ at $\geq 3\sigma$.
The results of the analysis \cite{Fogli:2011qn},
in which  $\Delta m^2_{21} \equiv \Delta m^2_{\odot}$
and $|\Delta m^2_{31}| \equiv |\Delta m^2_{\rm A}|$
were determined as well, are shown in Table \ref{tab:tabdata-1106}.

 Recently, the first data of the Daya Bay reactor
antineutrino experiment on $\theta_{13}$ was published
\cite{DBay0312}. The value of $\sin^22\theta_{13}$
was measured with a rather high precision
and was found to be different from zero
at $5.2\sigma$:
\begin{equation}
 \sin^22\theta_{13} = 0.092 \pm 0.016 \pm 0.005\,,~~
 0.04 \leq \sin^22\theta_{13} \leq 0.14\,,~3\sigma\,,
\label{DBayth13}
\end{equation}
%
where we have given also the $3\sigma$ interval
of allowed values of  $\sin^22\theta_{13}$.
Subsequently, the RENO experiment
reported a $4.9\sigma$ evidence
for a non-zero value of $\theta_{13}$ \cite{RENO0412}:
\begin{equation}
 \sin^22\theta_{13} = 0.113 \pm 0.013 \pm 0.019\,.
\label{RENOth13}
\end{equation}
%
The value of $\theta_{13}$ determined in the
RENO experiment is compatible with that measured
in the Daya Bay experiment. It is interesting to note
also that the mean value of $\sin^2\theta_{13}$
found in the global analysis of the neutrino
oscillation data in \cite{Fogli:2011qn}
differs very little from the mean values
found  in the Daya Bay and RENO experiments.

  The results on $\theta_{13}$ described above
will have far reaching implications
for the program of research in neutrino physics.
A relatively large value of
$\sin\theta_{13}\sim 0.15$
opens up the possibilities, in particular,
i) for searching for CP violation effects in
neutrino oscillations experiments
with high intensity accelerator
neutrino beams (like T2K, NO$\nu$A, etc.);
ii) for determining the sign of
$\Delta m^2_{32}$, and thus the type
of neutrino mass spectrum, which can be
with normal or inverted ordering (see, e.g.\ \cite{PDG10}),
in the long baseline neutrino oscillation experiments
at accelerators (NO$\nu$A, etc.),
in the experiments studying the oscillations
of atmospheric neutrinos (see, e.g.\ \cite{JBSP203}),
as well as in experiments with reactor antineutrinos
\cite{ReactNuHiera}. It has important implications
for the neutrinoless double beta ($\betabeta$-) decay
phenomenology in the case of neutrino mass spectrum
with normal ordering (NO) \cite{Pascoli:2007qh}.
A value of $\sin\theta_{13} \gtap 0.09$
is a necessary condition for a successful ``flavoured''
leptogenesis when the CP violation required for
the generation of the matter-antimatter asymmetry of the
Universe is provided entirely by the Dirac CP violating
phase in the neutrino mixing matrix \cite{PPRio106}.~\footnote{If indeed $\sin\theta_{13}\cong 0.15$
and the neutrino mass spectrum is with inverted
ordering (IO), further
important implications for ``flavoured''
leptogenesis are possible \cite{EMSTP08}.}
As was already discussed to some extent in the
literature and we will see further,
in certain specific cases
a value of  $\sin\theta_{13}\sim 0.15$
can have important implications also
for the phenomenology of the lepton flavour
violation (LFV) processes involving the
charged leptons in theories incorporating
one of the possible TeV scale see-saw
mechanisms of neutrino mass generation.

\begin{table} [t] 
\centering \caption{
\label{tab:tabdata-1106} The best-fit values and
$3\sigma$ allowed ranges of the 3-neutrino oscillation parameters,
derived from a global fit of the current neutrino oscillation data,
including the T2K and MINOS (but not the Daya Bay)
results (from ~\cite{Fogli:2011qn}).
The Daya Bay data \cite{DBay0312} on $\sin^2\theta_{13}$ is
given in the last line.
The values (values in brackets) of $\sin^2\theta_{12}$
are obtained using the ``old'' \cite{Schre85}
(``new'' \cite{Mention11})
reactor $\bar{\nu}_e$ fluxes in the analysis.}
\renewcommand{\arraystretch}{1.1}
\begin{tabular}{lcc}
\hline \hline
 Parameter  &  best-fit ($\pm 1\sigma$) & 3$\sigma$ \\ \hline
 $\Delta m^{2}_{\odot} \; [10^{-5}\eV^2]$   & 7.58$^{+0.22}_{-0.26}$ &
               6.99 - 8.18 \\
$ |\Delta m^{2}_{A}| \; [10^{-3}\eV^2]$ & 2.35$^{+0.12}_{-0.09}$  &
           2.06 - 2.67\\
$\sin^2\theta_{12}$  & 0.306$^{+0.018}_{-0.015}$
            & 0.259(0.265) - 0.359(0.364)\\
 $\sin^2\theta_{23}$  & 0.42$^{+0.08}_{-0.03}$ &  0.34 - 0.64 \\
$\sin^2\theta_{13}$ \cite{Fogli:2011qn}  &  0.021(0.025) $^{+0.007}_{-0.007}$  & 0.001(0.005) - 0.044(0.050) \\
$\sin^2\theta_{13}$ \cite{DBay0312}  &  0.0236 $\pm 0.0042$  & 0.010 - 0.036 \\
\hline\hline
\end{tabular}
\end{table}
%

  Despite the compelling evidence for the nonconservation
of the leptonic flavour in neutrino oscillations, all searches
for lepton flavour violation (LFV) in the charged lepton
sector have been so far negative. The best limits
follow from the non-observation of the LFV muon decays
$\mu^+\rightarrow e^+\gamma$ and $\mu^+\rightarrow e^+ e^- e^+$,
\begin{eqnarray}
\label{mutoegexp}
&~&{\rm BR}(\mu^+\rightarrow e^+\gamma)
< 2.4\times 10^{-12}~~\cite{Adam:2011ch}\,,\\
&~&{\rm BR}(\mu^+\rightarrow e^+ e^- e^+)< 1.0\times 10^{-12}~~
 \cite{Bellgardt:1987du}\,,
\label{muto3eexp}
\end{eqnarray}
%
and from the non-observation of conversion of muons
into electrons in Titanium,
\begin{equation}
{\rm CR}(\mu{\rm Ti}\rightarrow e{\rm Ti})\, <\,
4.3\times 10^{-12}~~~\cite{Dohmen:1993mp}\,.
\label{mu2eTi}
\end{equation}
%
  Besides, there are stringent constraints on the tau-muon
and tau-electron flavour violation from the non-observation of
LFV radiative tau decays ~\cite{Aubert:2009tk}:
\begin{eqnarray}
\label{tautomugexp}
&~&{\rm BR}(\tau\rightarrow \mu\gamma)<4.4\times 10^{-8}\,,\\
&~&{\rm BR}(\tau\rightarrow e \gamma)<3.3\times 10^{-8}\,.
\label{tautoegexp}
\end{eqnarray}
%

  The indicated stringent upper limits on the rates of the LFV
processes involving the charged leptons
lead to severe constraints on models of new physics which predict
new particles at the electroweak scale coupled to the charged leptons.
Indeed, the dipole operator which leads to the
process $\mu\rightarrow e\gamma$ has the form:
\begin{equation}
-{\cal L}=m_\mu \bar \mu (f^{\mu e}_{M1}
+ \, \gamma_5 f^{\mu e}_{E1})\sigma^{\nu\rho}
e F_{\nu\rho}+{\rm h.c.}
\end{equation}
%
where $f^{\mu e}_{M1}$ and  $f^{\mu e}_{E1}$ are, respectively,
the transition magnetic and electric dipole
moment form factors.
This operator is generated at the
quantum level through particles with masses
$\Lambda$ which couple to the
charged leptons, hence the form factors can be parameterised as
$f^{\mu e}=\frac{\theta^2_{\mu e}}{16\pi^2\Lambda^2}$,
$\theta_{\mu e}$ being a parameter which measures the strength
of the coupling of the new particles to
the electron and the muon. The present experimental limit
on ${\rm BR}(\mu\rightarrow e\gamma)$
sets the following upper limit on the two form factors:
$|f_{E1}^{\mu e}|,|f_{M1}^{\mu e}|\lesssim 10^{-12}\,{\rm GeV}^{-2}$.
The latter in turn translates into $\Lambda \gtrsim 20$ TeV if
$\theta_{\mu e}\sim 1/\sqrt{2}$,
or in $\theta_{\mu e}\lesssim 0.01$ if $\Lambda\sim 300$ GeV.
It is then apparent that experiments searching for
lepton flavour violation can probe models
of new physics which cannot be tested in collider experiments, either
because the new particles are not kinematically accessible
with the available collider energies,
or because the couplings of the new particles
to the Standard Model (SM) particles
are too feeble to produce the former with
rates necessary for their observation
given the luminosity of the present colliders.

  Low scale see-saw models are a particularly interesting class
of models of new physics which are severely constrained
by experiments searching for lepton flavour violation.
In this class of models the flavour structure of the couplings of the
new particles to the charged leptons is basically determined
by the requirement of reproducing the data on
the neutrino oscillation parameters
\cite{Raidal:2004vt,Shaposhnikov:2006nn,Kersten:2007vk,Gavela:2009cd,Ibarra:2011xn}.
Hence, the rates for the lepton flavour violating processes
in the charged lepton sector can be calculated in terms of a
few parameters, the predicted rates being possibly within the
reach of the future experiments searching
for lepton flavour violation,
even when the parameters of the model do not allow
production of the new particles with observable rates
at the LHC~\cite{Ibarra:2011xn}.

 The role of the experiments searching for lepton flavour
violation
to constrain low scale see-saw models will be
significantly strengthened in the next years.
Searches for $\mu-e$ conversion at the planned COMET experiment
at KEK~\cite{comet} and Mu2e experiment
at Fermilab~\cite{mu2e} aim to reach sensitivity to
$\rm{CR}(\mu\, {\rm Al} \to e\, {\rm Al})\approx  10^{-16}$,
while, in the longer run, the PRISM/PRIME experiment in KEK~\cite{PRIME}
and the project-X experiment in Fermilab~\cite{projectX}
are being designed to probe values of the $\mu-e$ conversion rate
on ${\rm Ti}$, which are by 2 orders of magnitude smaller,
$\rm{CR}(\mu\, {\rm Ti} \to e\, {\rm Ti})\approx 10^{-18}$~\cite{PRIME}.
If these experiments reach the projected sensitivity
without observing a signal,
the upper limits on the form factors $f_{M1}$, $f_{E1}$
will improve by two orders of magnitude.
There are also plans to perform a new search for the
$\mu^+\rightarrow e^+ e^- e^+$ decay \cite{muto3eNext},
which will probe values of the corresponding
branching ratio down to
${\rm BR}(\mu^+\rightarrow e^+ e^- e^+) \approx 10^{-15}$,
i.e., by 3 orders of magnitude smaller than
the best current upper limit eq.~(\ref{muto3eexp}).
Furthermore, searches for tau lepton flavour violation
at superB factories aim to reach a sensitivity to
${\rm BR}(\tau\rightarrow (\mu,e)\gamma)\approx 10^{-9}$
~\cite{Akeroyd:2004mj,Bona:2007qt}.

 In this paper we will study the constraints on low (TeV) scale
see-saw models of neutrino mass generation
from present and future
experiments searching for lepton flavour violation,
with especial emphasis on $\mu-e$ conversion in nuclei,
which is among all search strategies the one
with brightest perspectives.
The paper is organized as follows: in Sections \ref{TypeISec}, \ref{TypeIISec} and \ref{TypeIIISec}  we review the
main features of the three types of see-saw mechanisms.
We discuss for each scenario the predictions and experimental constraints on the relevant parameter space
arising from LFV processes. The results are summarized and discussed in the concluding
Section \ref{Conclusions}.

%
\section{TeV Scale Type I See-Saw Model}\label{TypeISec}
%
%
We consider in detail in this Section LFV processes emerging in type I see-saw
extensions of the SM~\cite{seesaw}.
 We denote the light and heavy Majorana
mass eigenstates of the
type I see-saw model as $\chi_i$ and $N_k$, respectively.
\footnote{We use the same notations as in
\cite{Ibarra:2011xn,Ibarra:2010xw}.}
The  charged and neutral current weak
interactions involving  the light Majorana neutrinos have the form:
\begin{eqnarray}
\label{nuCC}
\mathcal{L}_{CC}^\nu
&=& -\,\frac{g}{\sqrt{2}}\,
\bar{\ell}\,\gamma_{\alpha}\,\nu_{\ell L}\,W^{\alpha}\;
+\; {\rm h.c.}
=\, -\,\frac{g}{\sqrt{2}}\,
\bar{\ell}\,\gamma_{\alpha}\,
\left( (1+\eta)U \right)_{\ell i}\,\chi_{i L}\,W^{\alpha}\;
+\; {\rm h.c.}\,,\\
\label{nuNC}
\mathcal{L}_{NC}^\nu &=& -\, \frac{g}{2 c_{w}}\,
\overline{\nu_{\ell L}}\,\gamma_{\alpha}\,
\nu_{\ell L}\, Z^{\alpha}\;
= -\,\frac{g}{2 c_{w}}\,
\overline{\chi_{i L}}\,\gamma_{\alpha}\,
\left (U^\dagger(1+\eta+\eta^\dagger)U\right)_{ij}\,\chi_{j L}\,
Z^{\alpha}\,,
\end{eqnarray}
%
where $(1+\eta)U = U_{\rm PMNS}$ is the
Pontecorvo, Maki, Nakagawa, Sakata (PMNS)
neutrino mixing matrix \cite{BPont57,MNS62},
$U$ is a $3\times 3$ unitary
matrix which diagonalises the Majorana mass matrix
of the left-handed (LH) flavour neutrinos $\nu_{\ell L}$
(generated by the see-saw mechanism),
and the matrix $\eta$ characterises the deviations
from unitarity of the PMNS matrix.
The elements of  $U_{\rm PMNS}$ are determined
in experiments studying the oscillations of the
flavour neutrinos and antineutrinos, $\nu_\ell$ and $\bar{\nu}_\ell$,
$\ell=e,\mu,\tau$, at relatively low energies.
In these experiments the initial states of the
flavour neutrinos, produced in  weak processes,
are coherent superpositions of the states of
the light massive Majorana neutrino $\chi_i$ only.
The states of the heavy Majorana neutrino $N_j$ are
not present in the superpositions representing the
initial flavour neutrino states
and this leads to deviations
from unitarity of the PMNS matrix.

  The matrix $\eta$ can be expressed in terms of a matrix $RV$
whose elements  $(RV)_{\ell k}$ determine the strength
of the charged current (CC) and  neutral current (NC)
weak interaction couplings of the heavy
Majorana neutrinos $N_k$ to the $W^\pm$-boson and
the charged lepton $\ell$, and to the $Z^0$ boson and
the left-handed (LH) flavour neutrino $\nu_{\ell L}$,
$\ell=e,\mu,\tau$:
\begin{eqnarray}
 \mathcal{L}_{CC}^N &=& -\,\frac{g}{2\sqrt{2}}\,
\bar{\ell}\,\gamma_{\alpha}\,(RV)_{\ell k}(1 - \gamma_5)\,N_{k}\,W^{\alpha}\;
+\; {\rm h.c.}\,
\label{NCC},\\
 \mathcal{L}_{NC}^N &=& -\frac{g}{4 c_{w}}\,
\overline{\nu_{\ell L}}\,\gamma_{\alpha}\,(RV)_{\ell k}\,(1 - \gamma_5)\,N_{k}\,Z^{\alpha}\;
+\; {\rm h.c.}\,.
\label{NNC}
\end{eqnarray}
%
Here $V$ is the unitary matrix which diagonalises the
Majorana mass matrix of the heavy RH neutrinos and
the matrix $R$ is determined by (see \cite{Ibarra:2010xw})
$R^* \cong M_D\, M^{-1}_{N}$, $M_D$ and $M_N$ being the
neutrino Dirac and the RH neutrino Majorana mass matrices,
respectively, $|M_D| \ll |M_N|$. We have:
\begin{equation}
	\eta\;\equiv\;-\half R R^\dagger =
-\half (RV)(RV)^\dagger = \eta^\dagger\,.
\label{eta}
\end{equation}
%

  The elements of the matrices $RV$ and $\eta$
can be constrained by using the existing
neutrino oscillation data,
data on electroweak (EW) processes, etc.
\cite{Antusch:2008tz,Antusch:2006vwa}.
They should satisfy also the constraint
which is characteristic of the type I see-saw
mechanism under discussion:
\begin{equation}
|\sum_{k} (RV)^*_{\ell'k}\;M_k\, (RV)^{\dagger}_{k\ell}|
\cong |(m_{\nu})_{\ell'\ell}| \lesssim 1~{\rm eV}\,,
~\ell',\ell=e,\mu,\tau\,.
\label{VR1}
\end{equation}
%
Here $m_{\nu}$ is the Majorana mass matrix of the LH
flavour neutrinos generated by the see-saw mechanism.
The upper limit $|(m_{\nu})_{\ell'\ell}| \lesssim 1$ eV,
$\ell,\ell'=e,\mu,\tau$, follows from the existing data
on the neutrino masses and on the neutrino mixing
\cite{Merle:2006du}.
For the values of the masses $M_k$
of the heavy Majorana neutrinos $N_k$ of interest
for the present study, $M_{k}=\mathcal{O}(100-1000)$ GeV,
the simplest scheme in which the constraint
(\ref{VR1}) can be satisfied is \cite{Ibarra:2011xn}
the scheme with two heavy Majorana neutrinos
(see, e.g., \cite{3X2Models,Ibarra:2003up,PRST05}),
$N_1$ and $N_2$, which form a pseudo-Dirac neutrino
pair \cite{LW81,STPPD82}: $M_2 = M_1(1 + z)$, where $z \ll 1$,
which naturally arises in type I
see-saw models with a mildly
broken lepton number symmetry~\cite{Shaposhnikov:2006nn} and
in the inverse see-saw model \cite{Mohapatra:1986bd,LWDW83}.
In the scenario where the CC and NC couplings of $N_{1,2}$ are sizable,
the requirement of reproducing the correct low energy neutrino oscillation
parameters constrains significantly~\cite{Shaposhnikov:2006nn,Kersten:2007vk}
and in  certain cases determines the Yukawa couplings \cite{Raidal:2004vt,Gavela:2009cd,Ibarra:2011xn}. Correspondingly, the elements
 $(RV)_{\ell 1}$ and  $(RV)_{\ell 2}$ in  eqs. (\ref{NCC}) and  (\ref{NNC})
are also determined and in the case of interest take
the form \cite{Ibarra:2011xn}:
\begin{eqnarray}
\label{mixing-vs-y}
\left|\left(RV\right)_{\ell 1} \right|^{2}&=&
\frac{1}{2}\frac{y^{2} v^{2}}{M_{1}^{2}}\frac{m_{3}}{m_{2}+m_{3}}
    \left|U_{\ell 3}+i\sqrt{m_{2}/m_{3}}U_{\ell 2} \right|^{2}\,,
~~{\rm NH}\,,\\
\label{mixing-vs-yIH}
\left|\left(RV\right)_{\ell 1} \right|^{2}&=&
\frac{1}{2}\frac{y^{2} v^{2}}{M_{1}^{2}}\frac{m_{2}}{m_{1}+m_{2}}
    \left|U_{\ell 2}+i\sqrt{m_{1}/m_{2}}U_{\ell 1} \right|^{2}
\cong \;\frac{1}{4}\frac{y^{2} v^{2}}{M_{1}^{2}}
\left|U_{\ell 2}+iU_{\ell 1} \right|^{2}\,,
\,{\rm IH}\,,\\
(RV)_{\ell 2}&=&\pm i\, (RV)_{\ell 1}\sqrt{\frac{M_1}{M_2}}\,,
~\ell=e,\mu,\tau\,,
\label{rel0}
\end{eqnarray}
%
where $y$ represents the maximum eigenvalue of
the neutrino Yukawa matrix and $v\simeq174$ GeV.
In
eq. (\ref{mixing-vs-yIH}) we have used the fact that for
the IH spectrum one has $m_1 \cong m_2$.
Due to the relation (\ref{rel0})
between  $(RV)_{\ell 1}$ and  $(RV)_{\ell 2}$,
%
eq.~(\ref{VR1}) is automatically satisfied.

  Upper bounds on the couplings of RH neutrinos
with SM particles can be obtained from the low energy
electroweak precision data
on lepton number conserving processes like
$\pi\to\ell \overline{\nu}_{\ell}$, $Z\to \nu\overline{\nu}$
and other tree-level processes involving light neutrinos in the final
state \cite{Antusch:2008tz}.
These bounds read:
\begin{eqnarray}
 |(RV)_{e1}|^{2} & \lesssim & 2\times 10^{-3}
\label{e-bound}\,,\\
 |(RV)_{\mu 1}|^{2} &\lesssim & 0.8\times 10^{-3}
\label{mu-bound}\,,\\
|(RV)_{\tau 1}|^{2} & \lesssim & 2.6\times 10^{-3}
\label{tau-bound}\,.	
\end{eqnarray}
%

  Let us add that in the class of
type I see-saw models with
two heavy Majorana neutrinos we are considering
(see, e.g., \cite{3X2Models,Ibarra:2003up,PRST05}),
one of the three light (Majorana) neutrinos
is massless and hence the neutrino mass spectrum is
hierarchical. Two possible
types of hierarchical spectrum are
allowed by the current neutrino data
(see, e.g., \cite{PDG10}):
$i)$ normal hierarchical (NH),
$m_{1}=0$, $m_{2}=\sqrt{\dmsol}$ and $m_{3}=\sqrt{\dma}$,
where $\dmsol \equiv m^2_{2} -  m^2_{1} > 0$ and
$\dma \equiv m^2_{3} -  m^2_{1}$;
$ii)$ inverted hierarchical (IH),
$m_{3}=0$, $m_{2}=\sqrt{|\dma|}$ and
$m_{1}=\sqrt{|\dma| -\dmsol} \cong \sqrt{|\dma|}$,
where $\dmsol \equiv m^2_{2} -  m^2_{1} > 0$ and
$\dma =  m^2_{3} -  m^2_{2} < 0$.
In both cases we have: $\dmsol/|\dma| \cong 0.03 \ll 1$.

    The numerical results we will present further will be obtained
employing the standard parametrisation for the
unitary matrix $U$:
\begin{equation}
 U = V(\theta_{12},\theta_{23},\theta_{13},\delta)\,
 Q(\alpha_{21},\alpha_{31})\,.
 \label{UPMNS}
 \end{equation}
 %
 Here (see, e.g.,  \cite{PDG10})
\begin{equation} 
V = \left(
     \begin{array}{ccc}
       1 & 0 & 0 \\
       0 & c_{23} & s_{23} \\
       0 & -s_{23} & c_{23} \\
     \end{array}
   \right)\left(
            \begin{array}{ccc}
              c_{13} & 0 & s_{13}e^{-i\delta} \\
              0 & 1 & 0 \\
              -s_{13}e^{i\delta} & 0 & c_{13} \\
            \end{array}
          \right)\left(
                   \begin{array}{ccc}
                     c_{12} & s_{12} & 0 \\
                     -s_{12} & c_{12} & 0 \\
                     0 & 0 & 1 \\
                   \end{array}
                 \right)\,,
\label{V}
\end{equation}
%
where we have used the standard notation
$c_{ij} \equiv \cos\theta_{ij}$,
$s_{ij} \equiv \sin\theta_{ij}$, $\delta$ is the Dirac
CP violation (CPV) phase and the matrix $Q$ contains
the two Majorana CPV phases
\footnote{
In the case of the type II see-saw mechanism,
to be discussed in Section 3, we have $\eta = 0$
and thus the neutrino mixing matrix coincides
with $U$: $U_{\rm PMNS} = U$. We will employ the
parametrisation (\ref{UPMNS}) - (\ref{Q}) for $U$
also in that case.}
\cite{BHP80},
\begin{equation}
Q =  {\rm diag}(1, e^{i\alpha_{21}/2},e^{i\alpha_{31}/2})\,.
\label{Q}
\end{equation}
%
 We recall that $U_{\rm PMNS}= (1+\eta)U$. Thus, up to corrections
which depend on the elements of the matrix $\eta$ whose absolute
values, however, do not exceed approximately $5\times 10^{-3}$
\cite{Antusch:2008tz},
the values of the angles $\theta_{12}$, $\theta_{23}$ and $\theta_{13}$
coincide with the values of the solar neutrino, atmospheric neutrino and
the 1-3 (or ``reactor'') mixing angles, determined in the 3-neutrino mixing
analyses of the neutrino oscillation data and reported in
Table \ref{tab:tabdata-1106}.

 Given the neutrino masses and mixing angles,
the TeV scale type I see-saw scenario we are considering is
characterised by four parameters \cite{Ibarra:2011xn}:
the mass (scale) $M_1$, the Yukawa coupling $y$,
the parameter $z$ of the splitting between the masses
of the two heavy Majorana neutrinos and a phase $\omega$.
The mass $M_1$ and the Yukawa coupling $y$ can be determined,
in principle, from the measured rates of two lepton flavour
violating (LFV) processes, the
$\mu \to e \gamma$ decay and the $\mu - e$ conversion
in nuclei, for instance.
The mass splitting parameter $z$ and the phase $\omega$,
together with $M_1$ and $y$, enter, e.g., into the expression for
the rate of $(\beta\beta)_{0\nu}$-decay,
predicted by the model. The latter was discussed in detail in
\cite{Ibarra:2011xn}.

%
\mathversion{bold}
\subsection{The $\mu \to e \gamma$ Decay}
\mathversion{normal}
%

In this subsection we update briefly the discussion of the
limits on the parameters of the TeV scale type I see-saw model,
derived in \cite{Ibarra:2011xn} using the experimental upper
bound on the $\mu \to e \gamma$ decay rate obtained
in 1999 in the MEGA experiment \cite{Brooks:1999pu}.
After the publication of \cite{Ibarra:2011xn},
the MEG collaboration reported a new more stringent
upper bound on the $\mu \to e \gamma$ decay rate
\cite{Adam:2011ch} given in eq.~(\ref{mutoegexp}).
Such an update is also necessary in view of the
relatively large nonzero value of
the reactor angle $\theta_{13}$
measured in the Daya Bay and RENO experiments
\cite{DBay0312,RENO0412}
and reported in eqs. (\ref{DBayth13}) and (\ref{RENOth13}).
As was discussed in \cite{Ibarra:2011xn},
in particular, the rate of the
$\mu \to e \gamma$ decay
in the type I see-saw scheme considered can be
strongly suppressed for certain values of $\theta_{13}$.

  The $\mu\to e \gamma$ decay branching ratio in the
scenario under discussion is given by
\cite{Petcov:1976ff,Cheng:1980tp}:
\begin{eqnarray}
{\rm BR}(\mu\to e \gamma) =
\frac{\Gamma(\mu\to e \gamma)}{\Gamma(\mu\to e+\nu_{\mu}+\overline{\nu}_{e})}
&=&
\frac{3\alpha_{\rm em}}{32\pi}\,|T|^{2}\,,
\label{Bmutoeg1}
\end{eqnarray}
%
where $\alpha_{\rm em}$ is the fine structure constant
and \cite{Ibarra:2011xn}
\begin{equation}
|T|\;\cong\; \,
\frac{2 + z}{1 + z}\,
\left |(RV)_{\mu 1}^{*}\, (RV)_{e1}\right | \left| G(X) - G(0)\right|\,.
\label{T3b}
\end{equation}
%
In eqs. (\ref{Bmutoeg1}) and (\ref{T3b})
the loop integration function $G(x)$ has the form:
\begin{equation}
G(x)\;=\;\frac{10-43x+78 x^2 - 49 x^3 + 4 x^4 + 18 x^3 \log(x)}{3 (x - 1)^4}\,,
\label{G}
\end{equation}
%
where $X \equiv(M_{1}/M_{W})^{2}$. In deriving
the expression for the matrix element $T$,
eq.~(\ref{T3b}), we have
assumed that the difference between $M_1$ and $M_2$
is negligibly small and used $M_{1}\cong M_{2}$.
It is easy to verify that $G(x)$ is a monotonic
function which takes values in the interval $[4/3,10/3]$, with
$G(x)\cong\frac{10}{3}-x$ for $x\ll 1$.

  Using the expressions of $|(RV)_{\mu1}|^{2}$
and  $|(RV)_{e 1}|^{2}$ in terms of neutrino parameters,
eqs.~(\ref{mixing-vs-y}) and (\ref{mixing-vs-yIH}), we obtain
the $\mu\to e \gamma$ decay branching ratio for the NH and IH spectra:
\begin{eqnarray}
&&{\rm\bf NH:}\;\;\;{\rm BR}(\mu\to e \gamma)\cong\nonumber \\
&& \frac{3\alpha_{\rm em}}{32\pi}
\left(\frac{y^{2}v^{2}}{M_{1}^{2}}\frac{m_{3}}{m_{2}+m_{3}}\right)^{2}
\left|U_{\mu 3}+i\sqrt{\frac{m_{2}}{m_{3}}}U_{\mu 2} \right|^{2}
\left|U_{e 3}+i\sqrt{\frac{m_{2}}{m_{3}}}U_{e 2} \right|^{2}
\left[G(X)-G(0) \right]^{2}\,,
\label{meg-U-NH} \\\nonumber&&\\
&&{\rm\bf IH:}\;\;\;{\rm BR}(\mu\to e \gamma)\cong\nonumber\\
&& \frac{3\alpha_{\rm em}}{32\pi}
\left(\frac{y^{2}v^{2}}{M_{1}^{2}}\frac{1}{2}\right)^{2}
\left|U_{\mu 2}+iU_{\mu 1} \right|^{2}
\left|U_{e 2}+iU_{e 1} \right|^{2}
\left[G(X)-G(0) \right]^{2}\,.
\label{meg-U-IH}
\end{eqnarray}
%

  The data on the process
$\mu\to e \gamma$ set very stringent
constraints on the TeV scale type I see-saw mechanism.
The current upper bound on ${\rm BR}(\mu\to e \gamma)$
was obtained in the MEG experiment at PSI \cite{Adam:2011ch} and is
given in eq.~(\ref{mutoegexp}).
It is an improvement by a factor of 5 of the
previous best upper limit of the MEGA experiment,
published in 1999 \cite{Brooks:1999pu}.
The projected sensitivity of the
MEG experiment is
${\rm BR}(\mu \rightarrow e \gamma)\sim  10^{-13}$~\cite{Adam:2011ch}.
For $M_1 = 100~{\rm GeV}$ ($M_1 = 1$ TeV)
and $z \ll 1$ we get the following
upper limit on the product
$|(RV)_{\mu 1}^{*} (RV)_{e1}|$
of the heavy Majorana neutrino
couplings to the muon (electron)
and the $W^\pm$ boson
and to the $Z^0$ boson from the
the upper limit eq.~(\ref{mutoegexp}):
\begin{equation}
\left |(RV)_{\mu 1}^{*}\, (RV)_{e1}\right| < 0.8\times 10^{-4}\,
 (0.3\times 10^{-4})\,,
\label{T3}
\end{equation}
%
where we have used eqs.~(\ref{Bmutoeg1}) and (\ref{T3b}).
This can be recast as an upper bound on the
neutrino Yukawa coupling $y$. Taking, $e.g.$,
the best fit values of the solar and atmospheric
oscillation parameters given in Table \ref{tab:tabdata-1106}, we get:
\begin{eqnarray}
&& y\lesssim 0.035~(0.21)\,~{\rm for~NH~with~} M_1=100\,{\rm GeV}\,
(1000\,{\rm GeV})~{\rm and~}\sin\theta_{13}=0.1 \,,
\label{yupNH}\\
&& y\lesssim 0.025~(0.15)\,~{\rm for~IH~with~} M_1=100\,{\rm GeV}\,
(1000\,{\rm GeV})~{\rm and~}\sin\theta_{13}=0.1\,.
\label{yupIH}
\end{eqnarray}
%

 The constraints which follow from the
current MEG upper bound on ${\rm BR}(\mu\to e \gamma)$
will not be valid in the case of a cancellation
between the different terms in one of
the factors $|U_{\ell 3}+i\sqrt{m_{2}/m_{3}}U_{\ell 2}|^{2}$
and $|U_{\ell 2}+iU_{\ell 1}|^{2}$, $\ell=e,\mu$,
in the expressions (\ref{meg-U-NH}) and (\ref{meg-U-IH})
for ${\rm BR}(\mu\to e \gamma)$.
Employing the standard parametrisation of $U$,
eqs. (\ref{UPMNS}) - (\ref{Q}), one can show
that in the case of NH spectrum
we can have $|U_{e3}+i\sqrt{m_{2}/m_{3}}U_{e 2}| = 0$
if \cite{Ibarra:2011xn} (see also \cite{Raidal:2004vt})
$\sin(\delta+ (\alpha_{21}-\alpha_{31})/2) = 1$ and
$\tan\theta_{13}=(\dmsol/\dma)^{1/4}\sin\theta_{12}$.
Using the 3$\sigma$ allowed ranges of
$\dmsol$, $\dma$ and $\sin^2\theta_{12}$
given in Table \ref{tab:tabdata-1106},
we find that the second condition can be satisfied
provided $\sin^2\theta_{13} \gtap 0.04$,
which lies outside the $3\sigma$
range of allowed values of $\sin^2\theta_{13}$
found in the Daya Bay experiment \cite{DBay0312}
(see eq.~(\ref{DBayth13})).

\begin{figure}[t]
\begin{center}
\includegraphics[width=16cm,height=10cm]{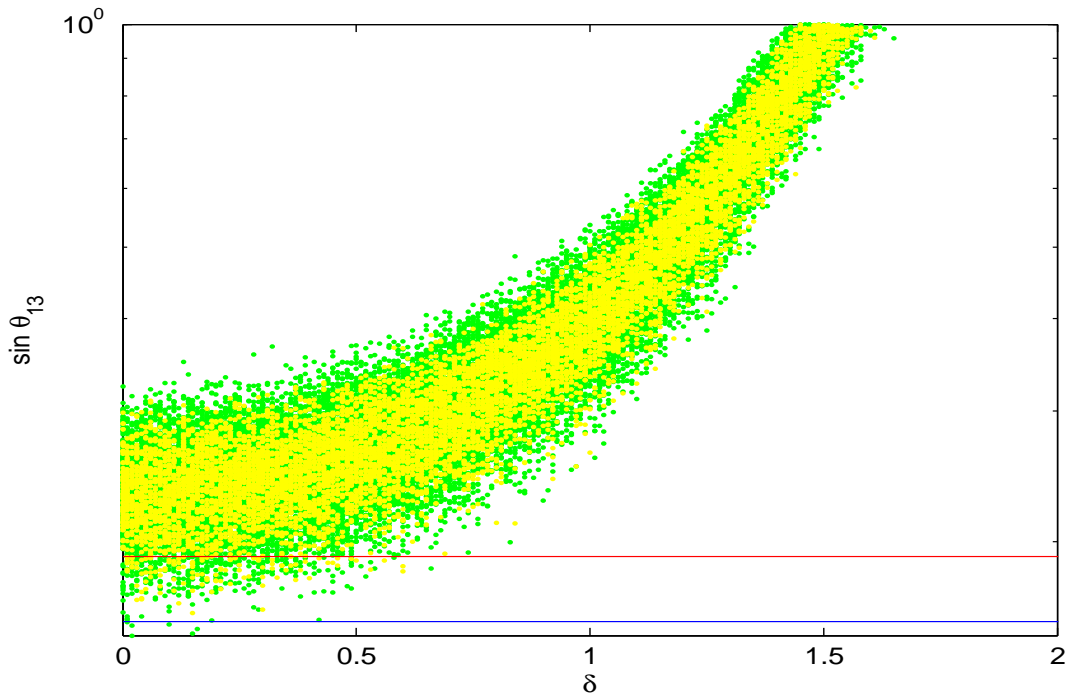}
\caption{Values of $\sin\theta_{13}$,
as a function of the phase $\delta$, which yield
a suppressed rate
of the process $\mu\to e\gamma$.
The values are obtained using eq. (\ref{s13min2}),
the 2$\sigma$ (3$\sigma$) intervals of allowed values of
$\sin^2\theta_{12}$ and $\sin^2\theta_{23}$,
yellow (green) points
(found with the ``new'' reactor $\bar{\nu}_e$ fluxes,
see Table \ref{tab:tabdata-1106})
and allowing $\delta$ to vary in the interval
[0,2$\pi$]. The red and blue horizontal lines correspond
to the $3\sigma$ upper limit
$\sin\theta_{13} = 0.191$ and the best fit value
$\sin\theta_{13} = 0.156$.
}
\label{th13suppr}
\end{center}
\end{figure}

 In the case of IH spectrum, the factor
$| U_{e 2} + i U_{e 1}|^{2}$  can be rather small for
$\sin (\alpha_{21}/2) = -1$: $| U_{e 2} + i U_{e 1}|^{2} =
c^2_{13}(1 - \sin2\theta_{12})
\cong 0.0765$,
where we have used the
best fit values of $\sin^2\theta_{12} = 0.306$ and
$\sin^2\theta_{13} = 0.0236$.
It is also possible to have a strong
suppression of the factor $|U_{\mu 2}+iU_{\mu 1}|^{2}$
\cite{Ibarra:2011xn}. Indeed, using the standard
parametrisation of the matrix $U$
it is not difficult to show that
for fixed values of the angles
$\theta_{12}$, $\theta_{23}$ and of the phases
$\alpha_{21}$ and $\delta$, $|U_{\mu 2}+iU_{\mu 1}|^{2}$
has a minimum for
\begin{eqnarray}
\sin\theta_{13} &=& \frac{c_{23}}{s_{23}}\,
\frac{\cos2\theta_{12} \cos\delta\sin\frac{\alpha_{21}}{2} -
\cos\frac{\alpha_{21}}{2}\sin\delta}
{1 + 2c_{12}\, s_{12}\, \sin\frac{\alpha_{21}}{2}}\,.
\label{s13min}
\end{eqnarray}
%
At the minimum we get:
\begin{eqnarray}
{\rm min}\left(|U_{\mu 2}+iU_{\mu 1}|^{2}\right)&=&
c^2_{23}\, \frac{\left(\cos\delta\cos\frac{\alpha_{21}}{2} +
\cos2\theta_{12}\sin\delta\sin\frac{\alpha_{21}}{2}\right)^{2}}
{1 + 2c_{12}\, s_{12}\,\sin\frac{\alpha_{21}}{2}}\,.
\label{Umu21IHmin}
\end{eqnarray}
%
Notice that, from the equation above, the $\mu \to e\gamma$ branching ratio is
highly suppressed if the Dirac and Majorana phases take CP conserving values, mainly: $\delta\simeq 0$ and $\alpha_{21}\simeq \pi$.
In this case, from eq.~(\ref{s13min}) we get the lower bound $\sin(\theta_{13})\gtrsim 0.13$, which is in agreement with the
Daya Bay measurement reported in Tab.~\ref{tab:tabdata-1106}. On the other hand, assuming CPV phases,
we still may have ${\rm min}(|U_{\mu 2}+iU_{\mu 1}|^{2})=0$, provided
$\theta_{12}$ and the Dirac and Majorana phases
$\delta$ and $\alpha_{21}$ satisfy the
following conditions:
$\cos\delta\cos(\alpha_{21}/2) +
\cos2\theta_{12}\sin\delta\sin(\alpha_{21}/2) = 0$
and ${\rm sgn}(\cos\delta\cos\frac{\alpha_{21}}{2})=
-{\rm sgn}(\sin\delta\sin\frac{\alpha_{21}}{2})$.
Taking $\cos\delta>0$ ($\cos\delta<0$) and using
$\tan\delta= - \tan(\alpha_{21}/2)/\cos2\theta_{12}$
in  eq.~(\ref{s13min}), we get the relation between
$s_{13}$, $\delta$ and $\cos2\theta_{12}$,
for which ${\rm min}(|U_{\mu 2}+iU_{\mu 1}|^{2})=0$:
\begin{equation}
\sin\theta_{13}\;=\,\frac{c_{23}}{s_{23}}\,
\frac{\sqrt{1 + \tan^{2}\delta}\,\cos2\theta_{12}}
{\sqrt{1 + \cos^22\theta_{12}\, \tan^{2}\delta} +
2c_{12}\, s_{12}\, {\rm sgn}(\cos\delta) }\,.
\label{s13min2}
\end{equation}
%
Using the 3$\sigma$ intervals of allowed values of
$\sin^2\theta_{12}$ and $\sin^2\theta_{23}$
(found with the ``new'' reactor $\bar{\nu}_e$ fluxes,
see Table \ref{tab:tabdata-1106})
and allowing $\delta$ to vary in the interval
[0,2$\pi$], we find that the values of $\sin\theta_{13}$
obtained using eq.~(\ref{s13min2}) lie in the interval
 $\sin\theta_{13} \gtap 0.11$.
As it follows from eq.~(\ref{DBayth13}), we have at $3\sigma$:
 $0.10\ltap \sin\theta_{13} \ltap 0.19$.
The values of $0.11 \ltap \sin\theta_{13} \ltap 0.19$
correspond to $0\leq \delta \ltap 0.7$.
These conclusions are illustrated in Fig. \ref{th13suppr}.
For $\sin\theta_{13}$ and $\delta$ lying in the
indicated intervals we can have  $|U_{\mu 2}+iU_{\mu 1}|^{2}=0$
and thus a strong suppression of
the $\mu \rightarrow e \gamma$ decay rate.
As we will see in subsections 2.2 and 2.3,
in the model we are considering,
the predicted $\mu-e$ conversion rate in a given nucleus
and $\mu\to 3e$ decay rate
are also proportional to
$|(RV)_{\mu 1}^{*}(RV)_{e1}|^{2}$,
as like the  $\mu \rightarrow e \gamma$
decay rate. This implies that in the case of
the TeV scale type I see-saw mechanism and
IH light neutrino mass spectrum,
if, e.g., ${\rm BR}(\mu\to e \gamma)$ is
strongly suppressed due to
$|U_{\mu 2}+iU_{\mu 1}|^{2} \cong 0$,
the $\mu-e$ conversion and the
$\mu \to 3e$ decay  rates
will also be strongly suppressed
\footnote{
Let us note that
in the case of IH spectrum we are
discussing actually one has
$|(RV)_{\mu 1}|^2 \propto
|U_{\mu 2}+i\sqrt{m_{1}/m_{2}}U_{\mu 1}|^{2}$
(see eq.~(\ref{mixing-vs-yIH})),
with $m_2 = \sqrt{|\Delta m^2_{\rm A}|}$ and
$m_1 = \sqrt{|\Delta m^2_{\rm A}| - \Delta m^2_{\odot}}
\cong  \sqrt{|\Delta m^2_{\rm A}|}
(1 - 0.5\Delta m^2_{\odot}/|\Delta m^2_{\rm A}|)$.
Therefore when $|U_{\mu 2}+iU_{\mu 1}| = 0$
we still have $|U_{\mu 2}+i\sqrt{m_{1}/m_{2}}U_{\mu 1}|^{2}\neq 0$.
However, in this case
$|U_{\mu 2}+i\sqrt{m_{1}/m_{2}}U_{\mu 1}|^{2} \cong
(\Delta m^2_{\odot}/(4|\Delta m^2_{\rm A}|))^2|U_{\mu 1}|^{2}
\ltap 1.7\times 10^{-5}$,
where we have used $\delta =0$ (which maximises $|U_{\mu 1}|^{2}$)
and the best fit values of the other neutrino oscillation parameters.
Thus, our conclusions about the suppression of
${\rm BR}(\mu\to e \gamma)$, the $\mu-e$ conversion and the
$\mu \to 3e$ decay  rates
are still valid.}.
The suppression under discussion
cannot hold if, for instance, it is
experimentally established that
$\delta$ is definitely bigger
than 1.0. That would be the case
if the existing indications
\cite{Fogli:2011qn} that $\cos\delta < 0$
receive unambiguous confirmation.
%

\begin{figure}
\begin{center}
\begin{tabular}{cc}
\includegraphics[width=7.5cm,height=6.5cm]{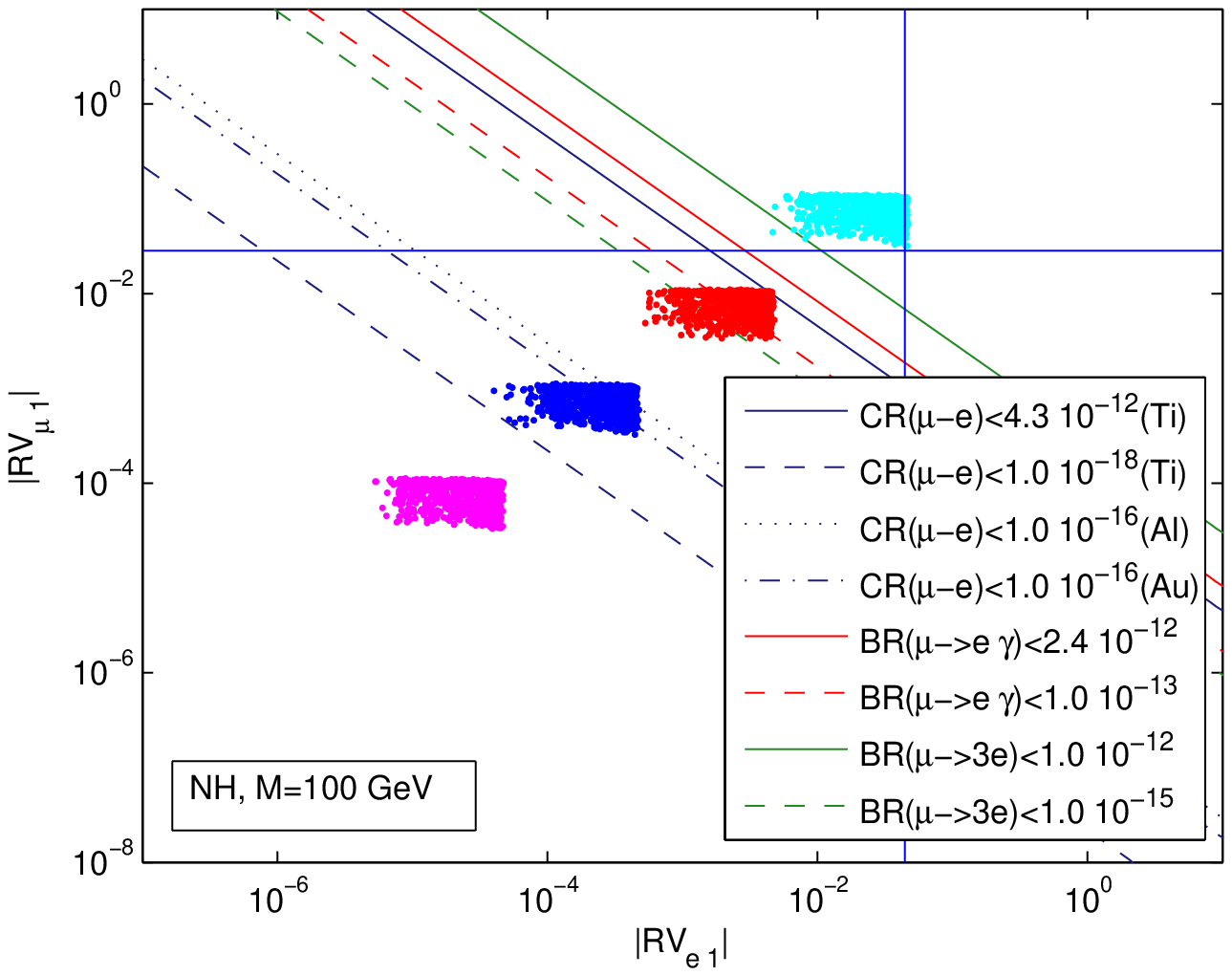} &
\includegraphics[width=7.5cm,height=6.5cm]{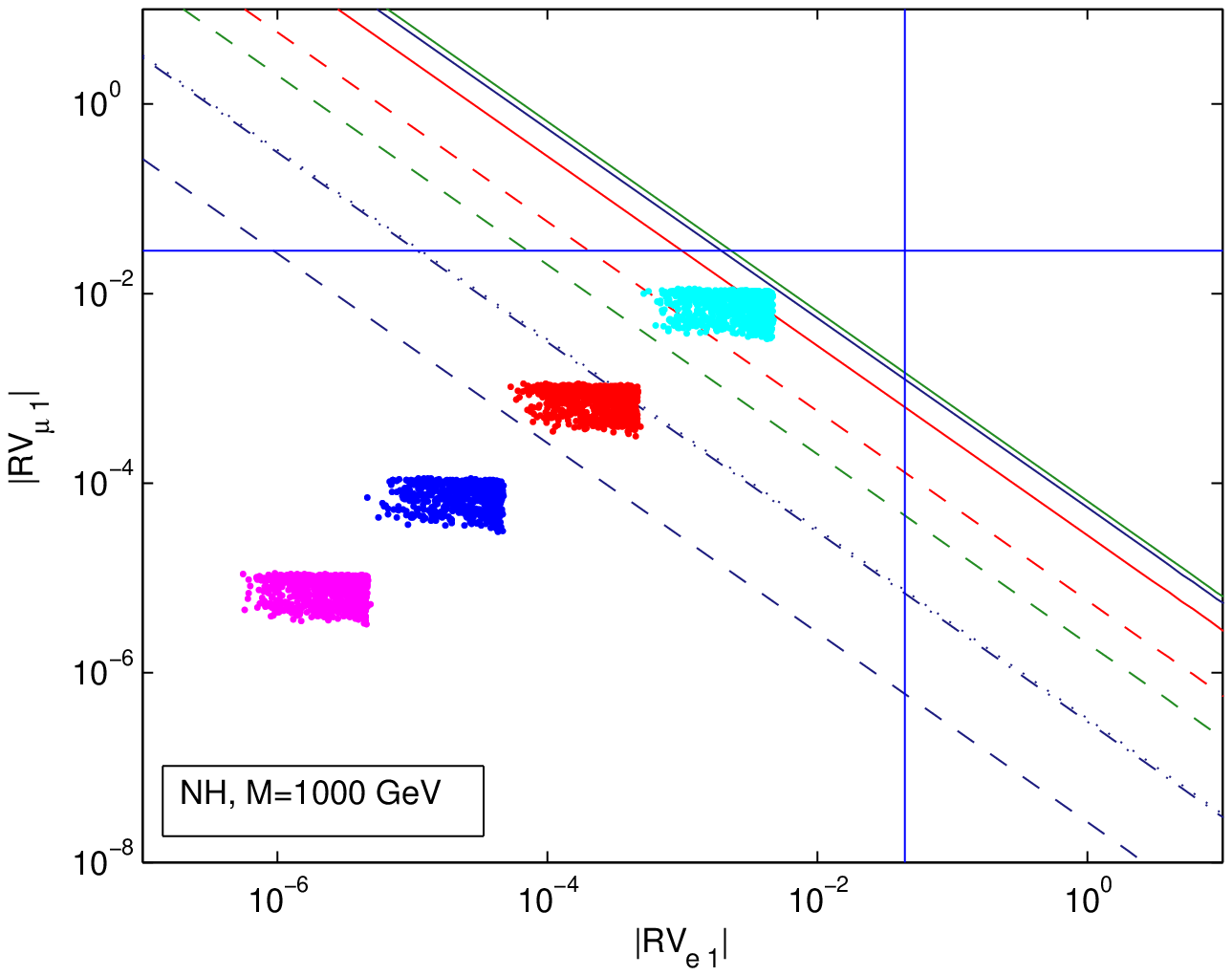}\\
\includegraphics[width=7.5cm,height=6.5cm]{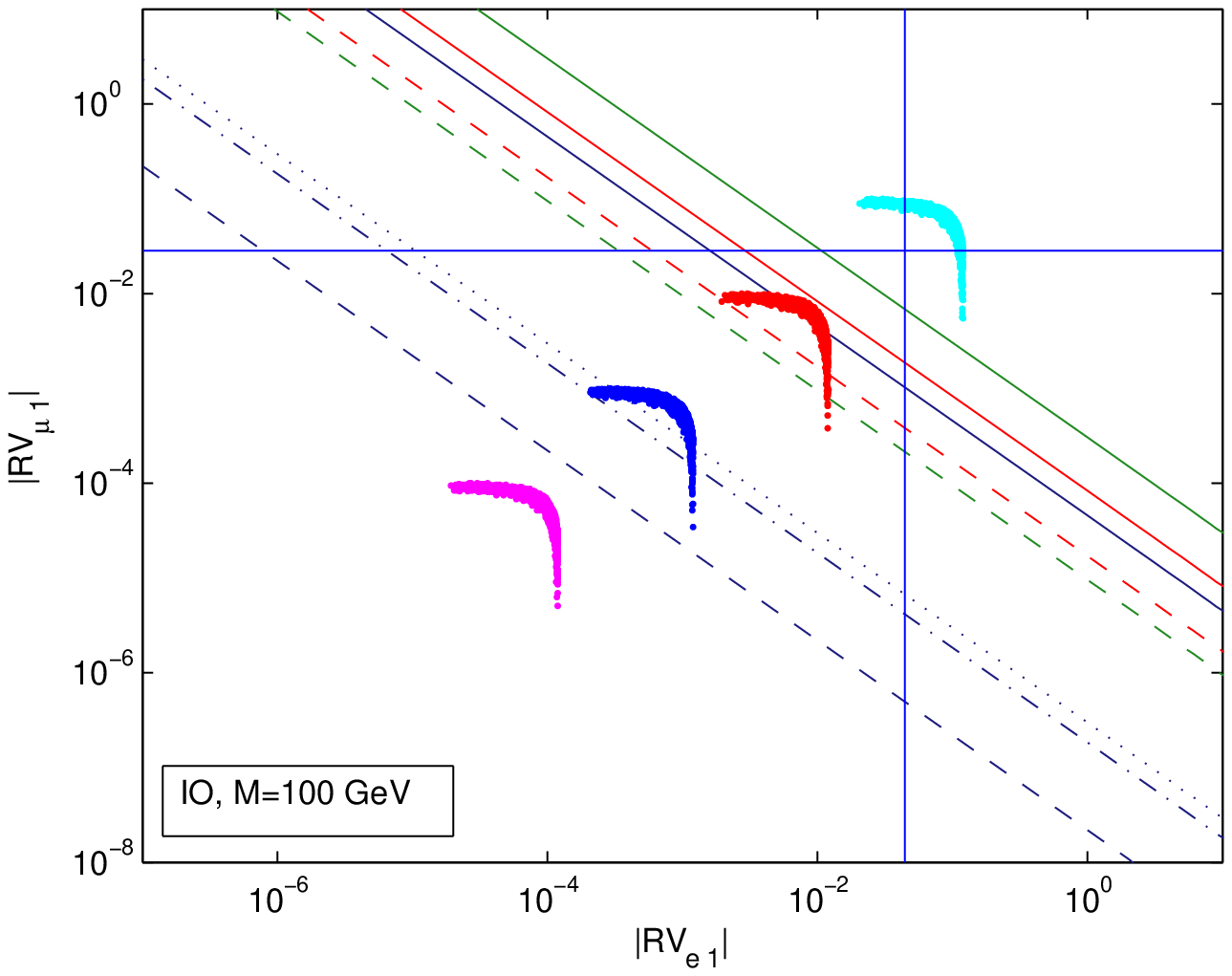} &
\includegraphics[width=7.5cm,height=6.5cm]{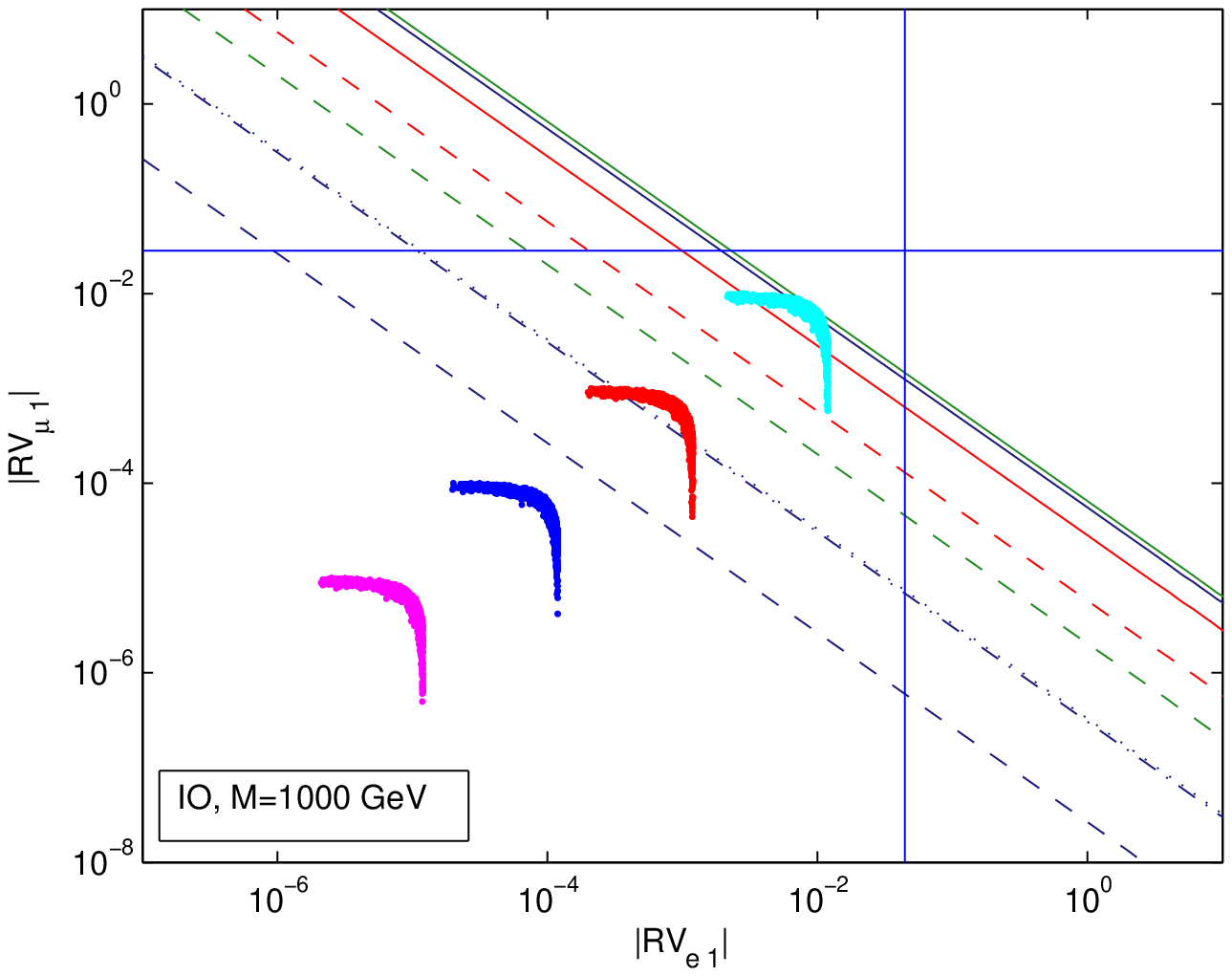}
\end{tabular}
\caption{Correlation between $|(RV)_{e1}|$ and
$|(RV)_{\mu1}|$ in the case of NH (upper panels)
and IH (lower panels) light neutrino mass spectrum, for
 $M_{1}=100\,(1000)$ GeV and,
$i)$ $y=0.0001$ (magenta points),
$ii)$ $y=0.001$ (blue points),
$iii)$ $y=0.01$ (red points) and
$iv)$ $y=0.1$ (cyan points).
The constraints from several LFV processes discussed in the text are shown.
}
\label{fig22}
\end{center}
\end{figure}

   The limits on the parameters  $|(RV)_{\mu 1}|$ and $|(RV)_{e1}|$,
implied by the electroweak precision data,
eqs. (\ref{e-bound}) - (\ref{tau-bound}),
and the upper bound on ${\rm BR}(\mu\to e \gamma)$, eq.~(\ref{mutoegexp}),
are illustrated in Fig.  \ref{fig22}. The results shown are
obtained for the best fit values of
$\sin\theta_{13} = 0.156$ and of the other
neutrino oscillation
parameters given in Table \ref{tab:tabdata-1106}.

%
\mathversion{bold}
\subsection{The $\mu - e$ Conversion in Nuclei}
\mathversion{normal}
%

We will discuss next the predictions of
the TeV scale type I see-saw extension of the SM
for the rate of the
$\mu-e$ conversion in nuclei, as well as
the experimental constraints
that can be imposed on this see-saw scenario
by the current and prospective  $\mu- e$ conversion data.
In the type I see-saw scenario
of interest, the $\mu-e$ conversion rate in a nucleus
$\mathcal{N}$ is very well approximated by
the expression \cite{Hisano96}:
\begin{equation}
{ \rm{CR}}(\mu\, \mathcal{N}\to e\, \mathcal{N})
\equiv \frac{\Gamma(\mu \,\mathcal{N}\to e\, \mathcal{N})}{\Gamma_{\rm capt}}=
 \frac{\alpha_{{\rm em}}^{5}}{2\,\pi^{4}\sin^4\theta_W}\,
\frac{Z_{eff}^{4}}{Z}\,\left| F(-m_{\mu}^{2}) \right|^{2}\,
 \frac{G_{F}^{2}m_{\mu}^{5}}{\Gamma_{\rm capt}}\,
\left| (RV)_{\mu 1}^{*}(RV)_{e1} \right|^{2}\,
\left| \mathcal{C}_{\mu e} \right|^{2} \; .\label{mu2e}
\end{equation}
In eq.~(\ref{mu2e}) $Z$
is the proton number of the nucleus $\mathcal{N}$,
$\theta_{W}$ is the weak mixing angle, $\sin^2\theta_{W} = 0.23$,
$F(-m^2_{\mu})$ is the nuclear form factor
at momentum transfer squared $q^2 = - m^2_{\mu}$,
$m_{\mu}$ being the muon mass, $Z_{eff}$ is an effective
atomic charge and $\Gamma_{\rm capt}$ is the experimentally known
total muon capture rate.
The loop integral factor $\mathcal{C}_{\mu e}$
receives contributions from $\gamma-$penguin,
$Z^{0}-$penguin and box type diagrams.
In the earlier version of the present article \cite{DinhIMP12}
we have used the expression for $|\mathcal{C}_{\mu e}|$
found in \cite{Hisano96} (in the notations of
ref. \cite{Buras10}) in a model with an active
heavy Majorana neutrino.
It was pointed out in \cite{Alonso12}, however,
that the result for  $|\mathcal{C}_{\mu e}|$ of
\cite{Hisano96} is not directly applicable to the
case of TeV scale type I see-saw model we are considering.
The authors of \cite{Alonso12} performed a detailed
calculation of $|\mathcal{C}_{\mu e}|$ in the model
of interest and obtained a new expression for
$|\mathcal{C}_{\mu e}|$.
We have performed an independent calculation of
the factor $|\mathcal{C}_{\mu e}|$ in the model
under discussion\footnote{The new results are published as an erratum to  \cite{DinhIMP12}.}. 
Our result for $|\mathcal{C}_{\mu e}|$ coincides with that
derived in \cite{Alonso12} and reads:
\begin{eqnarray}
\mathcal{C}_{\mu e} \cong
Z\left[2F_{u}^{\mu e}(X)+F_{d}^{\mu e}(X)\right]+N\left[F_{u}^{\mu e}(X)+2F_{d}^{\mu e}(X)\right]\,,~~~~~~~~~~~~~~\\
F_{q}^{\mu e}(X)=Q_q\sin^2{\theta_W}\left[F_\gamma(X)-F_z^{\mu e}(X)+G_{\gamma}(X)\right]
+\frac{1}{4}\left[2I_3F_z^{\mu e}(X)+F_B^{\mu e qq}(X)\right].
\label{Cmue}
\end{eqnarray}
Here
$N$ is the neutron number of the nucleus $\mathcal{N}$,
$X = M^2_{1}/M^2_{W}$, and
\begin{eqnarray}
\label{FG}
F_{\gamma}(x)&=&\frac{x(7x^2-x-12)}{12 (1-x)^3}-\frac{x^2(12-10x+x^2)}{6(1-x)^4}\log{x},  \,, \\ [0.30cm]
\label{GG}
G_{\gamma}(x)&=&-\frac{x(2x^2+5x-1)}{4 (1-x)^3}-\frac{3x^3}{2(1-x)^4}\log{x} \,, \\ [0.30cm]
\label{Fz}
F_z(x)&=&-\frac{5x}{2(1-x)}-\frac{5x^2}{2(1-x)^2}\log{x}\,, \\ [0.30cm]
\label{Gz}
G_z(x,y)&=&-\frac{1}{2(x-y)}\left[\frac{x^2(1-y)}{(1-x)}\log{x}-\frac{y^2(1-x)}{(1-y)}\log{y}\right]\,,
\end{eqnarray}
\begin{eqnarray}
\label{FBox}\nonumber
F_{Box}(x,y)&=&\frac{1}{x-y}\left\{(4+\frac{x y}{4})\left[\frac{1}{1-x}+\frac{x^2}{(1-x)^2}\log{x}-\frac{1}{1-y}
-\frac{y^2}{(1-y)^2}\log{y}\right]\right.\\
&&\left.-2xy\left[\frac{1}{1-x}+\frac{x}{(1-x)^2}\log{x}-\frac{1}{1-y}-\frac{y}{(1-y)^2}\log{y}\right]\right\}\,,\\
\label{FXBox}\nonumber
F_{XBox}(x,y)&=&-\frac{1}{x-y}\left\{(1+\frac{x y}{4})\left[\frac{1}{1-x}+\frac{x^2}{(1-x)^2}\log{x}-\frac{1}{1-y}-\frac{y^2}{(1-y)^2}\log{y}\right]\right.\\
&&\left.-2xy\left[\frac{1}{1-x}+\frac{x}{(1-x)^2}\log{x}-\frac{1}{1-y}-\frac{y}{(1-y)^2}\log{y}\right]\right\}\,,\\
&&F_z^{\mu e}(x)=F_z(x)+2G_z(0,x),~~F_{Box}^{\mu euu}(x)=F_{Box}(x,0)-F_{Box}(0,0)\,,\\
&&F_{Box}^{\mu edd}(x)=F_{XBox}(x,0)-F_{XBox}(0,0)\,.
\end{eqnarray}
\begin{figure}[t]
\begin{center}
\includegraphics[width=16cm,height=10cm]{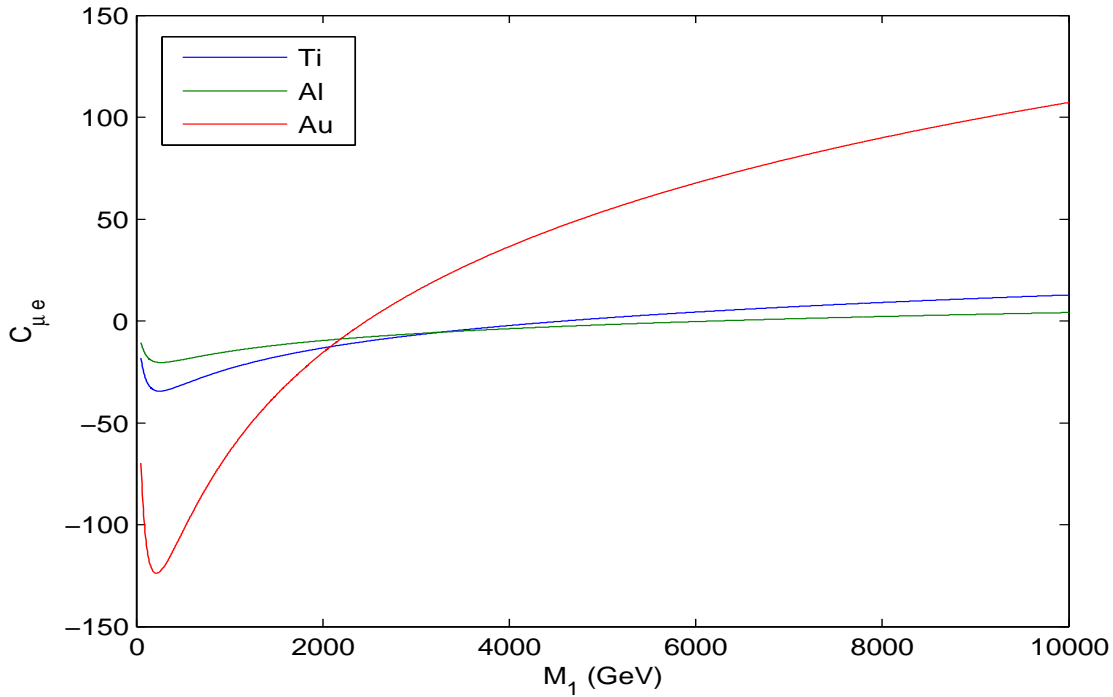}
\caption{The $\mu-e$ conversion loop integration factor $C_{\mu e}$
versus the see-saw mass scale $M_1$,
for three different nuclei: $i)$ $_{22}^{48}$Ti (blue line),
$ii)$ $_{13}^{27}$Al, (green line), and $iii)$ $_{79}^{197}$Au
(red line).}
\label{Cmue1}
\end{center}
\end{figure}
%

In what follows we will present results for three
nuclei which were used in the past, and are of interest
for possible future  $\mu-e$ conversion experiments:
$_{22}^{48}\text{Ti}$, $_{13}^{27}\text{Al}$ and
$_{79 }^{197}\text{Au}$. For these nuclei one has,
respectively: i) $Z_{eff}=17.6;~11.62~;~33.64$,
ii) $F(q^2=-m_{\mu}^2)\approx 0.54;~0.64;~0.20$, and
iii) $\Gamma_{\rm capt}=2.59;~0.69;~13.07\times 10^6\,{\rm sec^{-1}}$
\cite{PRIME}.

The dependence of the loop integration factor
$\mathcal{C}_{\mu e}$ on the see-saw mass scale $M_1$ for the three
nuclei of interest is shown in Fig. \ref{Cmue1}.
The first feature to notice is that
$|\mathcal{C}_{\mu e}|$ for $_{22}^{48}\text{Ti}$, $_{13}^{27}\text{Al}$ and
$_{79 }^{197}\text{Au}$ has maxima
$|\mathcal{C}_{\mu e}| = 34.4;~20.4;~124$ at $M_1=250;~267;~214$ GeV,
respectively. At $M_1=250$ GeV, $|\mathcal{C}_{\mu e}|$ for
$_{13}^{27}\text{Al}$ and $_{79 }^{197}\text{Au}$ takes the values
$|\mathcal{C}_{\mu e}(\text{Al})| \cong 20.4$
and $|\mathcal{C}_{\mu e}(\text{Au})| \cong 123.1$;
at  $M_1=267$ GeV, we have
$|\mathcal{C}_{\mu e}(\text{Ti})| \cong 34.4$
and $|\mathcal{C}_{\mu e}(\text{Au})| \cong 122.4$;
and finally, at $M_1=214$ GeV,
we find $|\mathcal{C}_{\mu e}(\text{Ti})| \cong 34.3$ and
$|\mathcal{C}_{\mu e}(\text{Al})| \cong 20.2$.
These maxima of $|\mathcal{C}_{\mu e}|$ give the biggest
enhancement factors for the conversion rate
when $M_1\leq 1000$ GeV. Beside the maxima, $|\mathcal{C}_{\mu e}|$
goes through zero at $M_1=4595;~6215;~2470$ GeV for
$_{22}^{48}\text{Ti}$, $_{13}^{27}\text{Al}$ and
$_{79 }^{197}\text{Au}$, respectively,
as was noticed also in \cite{Alonso12}.

   Qualitatively, the dependence of the factor
$|\mathcal{C}_{\mu e}|$ defined in eqs. (\ref{Cmue})-(\ref{FBox})
on $M_1$ exhibits the same features as the factor $|\mathcal{C}_{\mu e}|$
derived in \cite{Hisano96}, namely \cite{DinhIMP12},
at goes through zero at a certain value of $M_1 = M^0_1(\mathcal{N})$
which depends on the nucleus $\mathcal{N}$ and is a monotonically
increasing function of $M_1$ in the interval [50 GeV, $10^4$ GeV]
when $M_1$ decreases  (increases) starting from the value
$M_1 = M^0_1(\mathcal{N})$. The values of $M^0_1(\mathcal{N})$
at which  $|\mathcal{C}_{\mu e}|$ given in eqs. (\ref{Cmue})-(\ref{FBox})
and that obtained in \cite{Hisano96} are zero differ roughly by a
factor of 10 to 20, depending on the nucleus $\mathcal{N}$.

  For $M_1$ lying inside the interested
interval (100 - 1000) GeV,
the loop integration factor $|\mathcal{C}_{\mu e}|$ takes rather
large values for each of the three nuclei.
As our calculations show,
$|\mathcal{C}_{\mu e}|$ is not smaller than 23.4 for the Ti and 14.9
for the Al, while for the Au nucleus it exceeds 64.1.
Since the $\mu - e$ conversion rate is enhanced by the
factor $|\mathcal{C}_{\mu e}|^2$,
it is very sensitive to the product $|(RV)_{\mu 1}^{*}(RV)_{e1}|$
of CC couplings of the heavy Majorana neutrinos to
the electron and muon for the values of $M_1$ in
the interval of interest.

  The best experimental upper bound on the conversion
rate is \cite{Dohmen:1993mp}:
${\rm CR}(\mu\, {\rm Ti} \to e\, {\rm Ti}) \lesssim  4.3\times 10^{-12}$.
This bound implies a constraint on
$|(RV)_{\mu 1}^{*}(RV)_{e1}|$,
which is shown in Fig. \ref{fig22} for
$M_1 = 100;~1000$ GeV.
It is quite remarkable that, as
Fig. \ref{fig22} shows,
the constraint on the product of couplings
$|(RV)_{\mu 1}^{*}(RV)_{e1}|$ implied by the
best experimental upper limit on
${\rm CR}(\mu\, {\rm Ti} \to e\, {\rm Ti})$
is more stringent than the constraint following from the
best experimental upper limit on
${\rm BR}(\mu\to e \gamma)$ although the two experimental
upper limits are very similar quantitatively and
the expression for
${\rm CR}(\mu\, {\rm Ti} \to e\, {\rm Ti})$ has an additional
factor of $\alpha = 1/137$ with respect to
the expression for ${\rm BR}(\mu\to e \gamma)$.
\begin{figure}
\begin{center}
\begin{tabular}{cc}
\includegraphics[width=7.5cm,height=6.5cm]{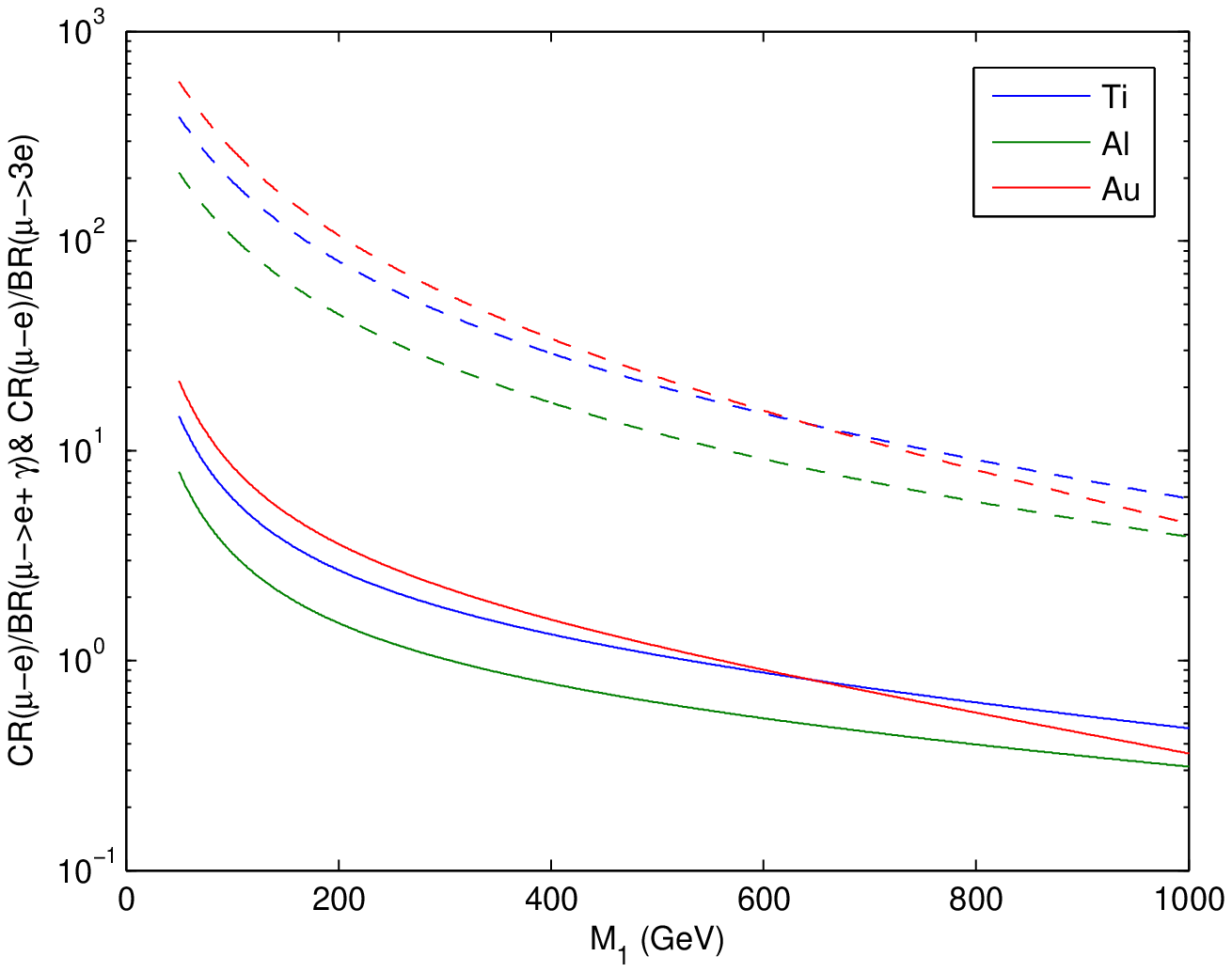} &
\includegraphics[width=7.5cm,height=6.5cm]{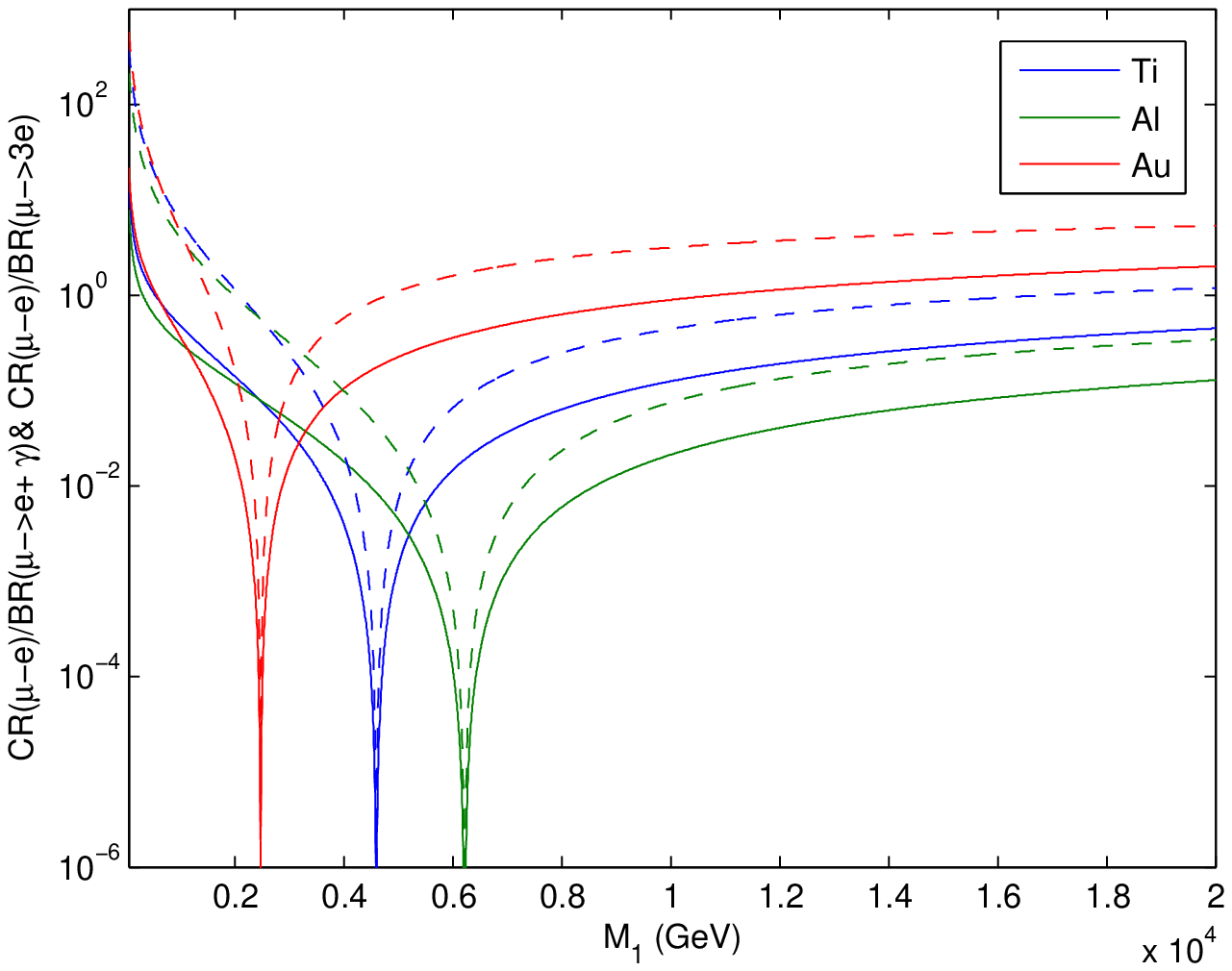}\\
\end{tabular}
\caption{The ratio of the $\mu-e$ relative conversion rate
and the branching ratio of the i) $\mu\to e\gamma$ decay (solid lines),
ii) $\mu\to 3e$ decay (dashed lines), versus
the type I see-saw mass scale $M_1$,
for three different nuclei:
$_{22}^{48}$Ti (blue lines), $_{13}^{27}$Al (green lines)
and $_{79}^{197}$Au (red lines).}
\label{CRvsBRs}
\end{center}
\end{figure}

  Future experimental searches for $\mu-e$ conversion in
$_{22}^{48}{\rm Ti}$ can reach the sensitivity of
${\rm CR}(\mu\, {\rm Ti} \to e\, {\rm Ti})
\approx 10^{-18}$ \cite{PRIME}. Therefore,
for values of $M_1$ outside the narrow intervals
quoted above for which the loop integration factor
 $|\mathcal{C}_{\mu e}|$ is strongly suppressed,
an upper bound on  the $\mu-e$ conversion ratio of
$\mathcal{O}(10^{-18})$ can
be translated into the following stringent constraint
on the heavy Majorana neutrino CC couplings to
the muon and electron:
\begin{equation}
\label{mueconv}
\left| (RV)_{\mu1}^{*} (RV)_{e1}\right| \;\lesssim\; 2.17\times10^{-8}~
(2.63\times10^{-8})\,~~{\rm for}~~M_{1}\approx100~(1000)~~{\rm GeV}\,.
\end{equation}

 As was noticed earlier, the two parameters of the type I
see-saw model considered, the mass scale $M_1$ and the Yukawa coupling $y$,
can be determined, in principle, from data on
${\rm BR}(\mu\to e \gamma)$ (or ${\rm BR}(\mu\to 3e)$)
and ${\rm CR}(\mu\, {\rm Ti} \to e\, {\rm Ti})$
if the two processes will be observed. Actually,
the ratio  of the rates of $\mu-e$  conversion in any given nucleus
$\mathcal{N}$, ${\rm CR}(\mu\, \mathcal{N}\to e\, \mathcal{N})$,
and of the $\mu\rightarrow e\gamma$ decay, depends only
on the mass (scale) $M_1$ and can be used, in principle,
to determine the latter.
In the case of $\mu-e$ conversion
on titanium, for instance, we find:
\begin{equation}
R\left(\frac{\mu-e}{\mu \to e\gamma}\right) \equiv \frac{{\rm CR}(\mu\, {\rm Ti} \to e\,{\rm Ti})}{{\rm BR}(\mu\rightarrow e\gamma)}
\;\approx\; 5.95~(0.48)\,~~
{\rm for}~~M_{1}\approx 100~(1000)~~{\rm GeV}\,.
\end{equation}

The correlation between
${\rm CR}(\mu\, \mathcal{N}\to e\, \mathcal{N})$
and ${\rm BR}(\mu\rightarrow e\gamma)$  in the model considered
is illustrated in Fig. \ref{CRvsBRs}. The type I see-saw  mass
scale $M_1$ would be
uniquely determined if $\mu - e$ conversion is observed in
two different nuclei or if, e.g., the $\mu\rightarrow e\gamma$
decay or $\mu - e$ conversion in a given nucleus
is observed and it is experimentally
established that $R(\frac{\mu-e}{\mu \to e\gamma})\ltap 10^{-3}$.
In the latter case $M_1$ could be determined
with a relatively high precision.
Furthermore, as Fig.  \ref{CRvsBRs} indicates, if
the RH neutrino mass $M_1$ lies in the interval $(50-1000)$ GeV,
$M_1$ would be uniquely determined provided
$R(\frac{\mu-e}{\mu \to e\gamma})$ is measured with a
sufficiently high precision.

   We note also that the correlation between
 ${\rm CR}(\mu\, \mathcal{N}\to e\, \mathcal{N})$
and ${\rm BR}(\mu\rightarrow e\gamma)$ in the
type I see-saw model considered is qualitatively
and quantitatively very different from the correlation
in models where the $\mu-e$ conversion is
dominated by the photon penguin diagram, e.g.,
the supersymmetric high-scale see-saw model which
predicts  approximately \cite{Hisano:2001qz}
 ${\rm CR}(\mu\, {\rm Ti} \to e\,{\rm Ti})
\approx 5\times 10^{-3}\,{\rm BR}(\mu\rightarrow e\gamma)$.

%
\mathversion{bold}
\subsection{The $\mu \to 3e$ decay}
\mathversion{normal}
%
  The $\mu \to 3e$ decay branching ratio has
been calculated in \cite{Pilaftsis} in a
type I seesaw mechanism of neutrino mass generation
with arbitrary fixed number of heavy RH neutrinos.
After recalculating the form factors and neglecting
the effects of mass difference between $N_1$ and $N_2$,
we find in the model of interest to leading order in the
small parameters $|(RV)_{l 1}|$:\footnote{The new results are published as an erratum to \cite{DinhIMP12}.}
\begin{eqnarray}
\label{mu3e}
{\rm BR}(\mu \to 3 e)
&=& \frac{\alpha_{em}^2}{16\pi^2\sin^4{\theta_W}}\left| (RV)_{\mu1}^{*} (RV)_{e1}\right|^2\,
\left |C_{\mu3e}(x)\right |^2,\\
\nonumber
\left |C_{\mu3e}(x)\right |^2 &=& 2\left|\frac{1}{2}F_B^{\mu3e}+F_z^{\mu3e}-2\sin^2{\theta_W}(F_z^{\mu3e}-F_{\gamma})\right|^2
+4\sin^4{\theta_W}\left|F_z^{\mu3e}-F_{\gamma}\right|^2\\
\nonumber
&&+16\sin^2{\theta_W}
\left[(F_z^{\mu3e}+\frac{1}{2}F_B^{\mu3e})G_{\gamma}\right]
- 48\sin^4{\theta_W}\left[(F_z^{\mu3e}-F_\gamma^{\mu3e})G_{\gamma}\right]\\
&&+32\sin^4{\theta_W}|G_\gamma|^2\left(\log{\frac{m_\mu^2}{m_e^2}}-\frac{11}{4}\right)\,,
\label{cm3e}
\end{eqnarray}
%
where $F_\gamma(x)$, $G_\gamma(x)$, $F_z(x)$, $G_z(x,y)$, $F_{XBox}(x,y)$ are defined in (\ref{FG}),
(\ref{GG}), (\ref{Fz}), (\ref{Gz}), (\ref{FXBox}), and
\begin{eqnarray}
F_z^{\mu3e}(x)=F_z(x)+2G_z(0,x),\,\,\,F_B^{\mu3e}(x)=-2(F_{XBox}(0,x)-F_{XBox}(0,0)).
\end{eqnarray}
\begin{figure}[t]
\begin{center}
\includegraphics[width=16cm,height=10cm]{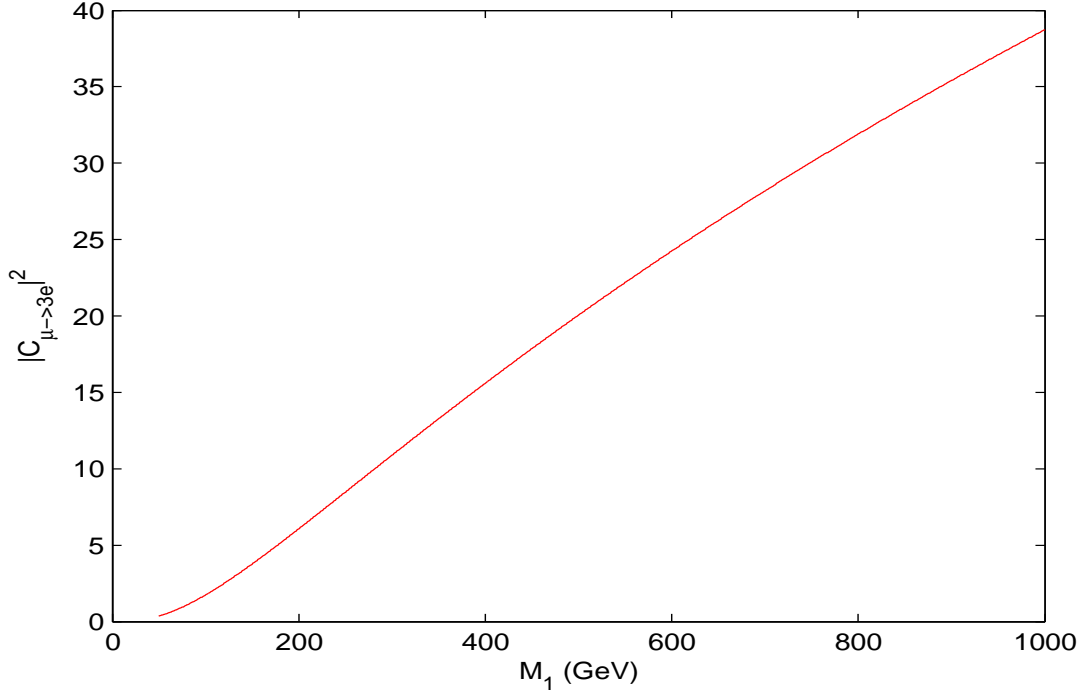}
\caption{The $\mu \to 3e$ decay rate
factor $|C_{\mu 3e}|^2$  as a function of the
see-saw mass scale $M_1$.
}
\label{Cmu3e}
\end{center}
\end{figure}
%

The dependence of the
$\mu \to 3e$ decay rate factor
$|C_{\mu 3e}|^2$ on the type I see-saw mass scale $M_1$
is shown in Fig. \ref{Cmu3e}. At $M_1 = 100~(1000)$ GeV we have:
$|C_{\mu 3e}|^2 \cong 1.75~(38.73)$, i.e.,
$|C_{\mu 3e}|^2$ increases by a factor of 22 when
$M_1$ changes from 100 GeV to 1000 GeV.
Using the quoted  values of $|C_{\mu 3e}|^2$
we get the following constraint from
the current limit on ${\rm BR}(\mu \to 3e)$, eq.~(\ref{muto3eexp}):
\begin{equation}
\label{mu3econstr}
\left| (RV)_{\mu1}^{*} (RV)_{e1}\right| \;\lesssim\; 3.01\times 10^{-4}~
(6.39\times10^{-6})\,~~{\rm for}~~M_{1} = 100~(1000)~~{\rm GeV}\,.
\end{equation}
Thus, for  $M_1 = 100$ GeV
the constraint on  $|(RV)_{\mu1}^{*} (RV)_{e1}|$
obtained using the current experimental upper limit on
${\rm BR}(\mu \to 3e)$ is by a factor of 3.90 less stringent
than that obtained from the current upper limit on
${\rm BR}(\mu \to e \gamma)$ (see eq.~(\ref{T3})),
while for  $M_1 = 1000$ GeV it is by a factor
of 4.7 more stringent. However, for the two values of
$M_1$ considered, the upper limit on
$|(RV)_{\mu1}^{*} (RV)_{e1}|$ from the current
experimental bound on the $\mu - e$ conversion rate,
eq.~(\ref{mu2eTi}), is the most stringent.
This is clearly seen in Fig. \ref{fig22}.
It follows also from Fig. \ref{fig22} that
an experiment sensitive to a $\mu-e$ conversion rate
${\rm CR}(\mu\, {\rm Al} \to e\, {\rm Al})\approx  10^{-16}$,
will probe smaller values of the
product of couplings $|(RV)_{\mu1}^{*} (RV)_{e1}|$
than an experiment sensitive to
${\rm BR}(\mu \to 3e) = 10^{-15}$.

  In Fig. \ref{CRvsBRs} we show the
correlation between  ${\rm CR}(\mu\, \mathcal{N}\to e\, \mathcal{N})$
and ${\rm BR}(\mu\rightarrow 3e)$ in the TeV scale see-saw
model considered. As it follows from Fig. \ref{CRvsBRs},
the observation of the $\mu\rightarrow 3e$ decay
or of the $\mu - e$ conversion in a given nucleus,
combined with data on the ratio
${\rm CR}(\mu\, \mathcal{N}\to e\, \mathcal{N})/{\rm BR}(\mu\rightarrow 3e)$
would lead either to a unique determination of
the type I see-saw scale $M_1$, or to two values,
or else to a relatively narrow interval of values,
of $M_1$ compatible with the data.
One can get the same type of information on the scale $M_1$
from data on the ratio  ${\rm BR}(\mu\rightarrow 3e)/{\rm BR}(\mu\rightarrow e \gamma)$,
provided at least one of the two decays
$\mu\rightarrow e \gamma$ and $\mu\rightarrow 3e$
is observed.

It should be added finally that for  $M_1\gtap 100$ GeV we have:
${\rm BR}(\mu\rightarrow 3e)/{\rm BR}(\mu\rightarrow e \gamma) \gtap 0.031$.
Thus, if it is experimentally established that
${\rm BR}(\mu\rightarrow 3e)/{\rm BR}(\mu\rightarrow e \gamma)$
is definitely smaller than the quoted lower bound,
the model considered with $M_1\gtap 100$ GeV will be
ruled out. Such a result would be consistent also just
with a see-saw scale $M_1 < 100$ GeV.

%
\section{TeV Scale Type II See-Saw Model}\label{TypeIISec}
 We will consider in this section the type II see-saw  \cite{typeII}
extension of the SM for the generation of the light neutrino masses.
In its minimal formulation it includes one additional
$SU(2)_{L}$ triplet Higgs field  $\Delta$, which has
weak hypercharge $Y_W=2$:
\begin{equation}
	\Delta\;=\;\left(
			\begin{array}{cc}
				\Delta^{+}/\sqrt{2} & \Delta^{++} \\
				\Delta^{0}	& -\Delta^{+}/\sqrt{2}
			\end{array}\right)\,.
\end{equation}
%
The Lagrangian of the type II see-saw scenario,
which is sometimes called also the ``Higgs Triplet Model'' (HTM),
reads
~\footnote{We do not give here, for simplicity,
all the quadratic and quartic
terms present in the scalar potential
(see, $e.g.$, \cite{Akeroyd:2009nu}).
}:
\begin{eqnarray}
	\mathcal{L}^{\rm II}_{\rm seesaw} &=& -M_{\Delta}^{2}\,{\rm Tr}\left(\Delta^{\dagger}\Delta\right)
		-\left(h_{\ell\ell^{\prime}}\,\overline{\psi^{C}}_{\ell L}\,i\tau_{2}\,\Delta\,\psi_{\ell^{\prime}L}
		\,+\,\mu_{\Delta}\, H^{T}\,i\tau_{2}\,\Delta^{\dagger}\,H\,+\,{\rm h.c.}\right)\,,
\label{LtypeII}
\end{eqnarray}
%
where $(\psi_{\ell L})^T \equiv (\nu^T_{\ell L}~~\ell^T_{L})$,
$\overline{\psi^{C}}_{\ell L}
\equiv ( -\,\nu^T_{\ell L}C^{-1}~~-\,\ell^T_{L}C^{-1})$,
and $H$ are, respectively, the SM lepton and Higgs doublets,
$C$ being the charge conjugation matrix, and
$\mu_{\Delta}$ is a real parameter characterising the
soft explicit breaking of the total lepton charge
conservation. We are interested in the low energy see-saw scenario,
where the new physics scale $M_{\Delta}$
associated with the mass of $\Delta$ takes values
$100~{\rm GeV}\lesssim M_{\Delta} \lesssim 1~{\rm TeV}$,
which, in principle, can be probed by LHC~\cite{colliders}.

 The flavour structure of the Yukawa coupling matrix $h$ and
the size of the lepton charge soft breaking parameter $\mu_{\Delta}$
are related to the light neutrino mass matrix $m_{\nu}$,
which is generated when the neutral component of $\Delta$
develops a ``small'' vev  $v_{\Delta} \propto \mu_{\Delta}$ .
Indeed, setting
$\Delta^{0} = v_{\Delta}$ and $H^T  = (0~~v)^{T}$
with $v\simeq 174$ GeV,
from  Lagrangian (\ref{LtypeII}) one obtains:
\begin{equation}
\left(m_{\nu}\right)_{\ell\ell^{\prime}}\, \equiv m_{\ell\ell^{\prime}}\,
\simeq\;2\,h_{\ell\ell^{\prime}}\,
v_{\Delta}\;.
\label{mnuII}
\end{equation}
%
The matrix of Yukawa couplings $h_{\ell\ell^\prime}$ is directly
related to the PMNS neutrino mixing matrix $U_{\rm PMNS}\equiv U$,
which is unitary in this case:
\begin{equation}
h_{\ell\ell^\prime}\;\equiv\; \frac{1}{2v_\Delta}\left(U^*\,
\diag(m_1,m_2,m_3)\,U^\dagger\right)_{\ell\ell^\prime}\,.
\label{hU}
\end{equation}
%
An upper limit on $v_\Delta$ can be obtained from
considering its effect on the parameter $\rho=M^2_W/M_Z^2\cos^2\theta_W$.
In the SM, $\rho=1$ at tree-level, while in the HTM one has
\begin{equation}
\rho\equiv 1+\delta\rho={1+2x^2\over 1+4x^2}\,,~~~x \equiv v_\Delta/v.
\label{deltarho}
\end{equation}
%
The measurement $\rho\approx 1$ leads to the bound
$v_\Delta/v\lesssim 0.03$, or  $v_\Delta<5$~GeV
(see, e.g., \cite{vDrho}).

 As we will see, the amplitudes of the LFV processes
 $\mu\rightarrow e \gamma$, $\mu\rightarrow 3e$ and
 $\mu + \mathcal{N}\to e + \mathcal{N}$ in the model
under discussion are proportional, to leading order,
to a product of 2 elements of the Yukawa coupling matrix $h$.
This implies that in order for the rates of the indicated
LFV processes to be close to the existing upper limits and
within the sensitivity of the ongoing MEG and the planned
future experiments for $M_{\Delta} \sim (100 - 1000)$ GeV,
the Higgs triplet vacuum expectation value $v_{\Delta}$
must be relatively small, roughly  $v_{\Delta} \sim (1 - 100)$ eV.
In the case of $M_{\Delta} \sim v = 174$ GeV we have
$v_{\Delta} \cong \mu_{\Delta}$, while if
$M^2_{\Delta} >> v^2$, then
$v_{\Delta} \cong \mu_{\Delta} v^2/(2M^2_{\Delta})$
(see, e.g., \cite{Akeroyd:2009nu,vDrho})
with $v^2/(2M^2_{\Delta}) \cong 0.015$ for $M_{\Delta} = 1000$ GeV.
Thus, in both cases a relatively small value of $v_{\Delta}$
implies that $\mu_{\Delta}$ has also to be small.
A nonzero but relatively small value of  $\mu_{\Delta}$
can be generated, e.g.,  at higher orders in perturbation
theory~\cite{Chun:2003ej} or in the context of theories with
extra dimensions~(see, e.g., \cite{Chen:2005mz}).

  The physical singly-charged Higgs scalar field
 practically coincides with the triplet scalar
field $\Delta^{+}$, the admixture of the doublet charged
scalar field being suppressed by the
factor $v_{\Delta}/v$. The singly- and doubly-  charged
Higgs scalars $\Delta^{+}$ and $\Delta^{++}$ have,
in general, different masses \cite{Chun:2003ej,HiggsMass}:
$m_{\Delta^{+}}\neq m_{\Delta^{++}}$.
Both situations $m_{\Delta^{+}} >  m_{\Delta^{++}}$ and
$m_{\Delta^{+}} < m_{\Delta^{++}}$ are possible.
In some cases, for simplicity,
we will present numerical results for
$m_{\Delta^{+}} \cong m_{\Delta^{++}} \equiv M_{\Delta}$, but
one must keep in mind that $m_{\Delta^{+}}$ and $m_{\Delta^{++}}$
can have different values.

  In the mass eigenstate basis, the effective charged lepton
flavour changing operators arise at one-loop order from the
exchange of the singly- and doubly-charged
physical Higgs scalar fields.
The corresponding effective low energy LFV Lagrangian, which contributes
to the $\mu-e$ transition processes we are interested in,
can be written in the form:
\begin{eqnarray}
\mathcal{L}^{eff} &=& -\,4\,\frac{e\,G_F}{\sqrt{2}}\,
\left(m_\mu\,A_R\,
\overline{e}\,\sigma^{\alpha\beta}\,P_R\,\mu\,F_{\beta\alpha}\,
+\,{\rm h.c.}\right) \nonumber \\
&&-\,\frac{e^2\,G_F}{\sqrt{2}}\,
\left(A_L(-m_\mu^2)\,\overline{e}\,\gamma^\alpha\,P_L\,\mu\,\,
\sum_{Q=u,d}\, q_{Q}\, \overline{Q}\,\gamma_\alpha\,Q\,
+\,{\rm h.c.} \right)\,,
\label{LefftypeII}
\end{eqnarray}
%
where $e$ is the proton charge, and
$q_{u}= 2/3$ and $q_{d}=-1/3$
are the electric charges of the up and
down quarks (in units of the proton charge).
We obtain for the form factors $A_{R,L}$:
\begin{eqnarray}
	A_R &=& -\,\frac{1}{\sqrt{2}\,G_F}
\frac{\left(h^\dagger h\right)_{e\mu}}{48\pi^2}\,
\left(\frac{1}{8\,m_{\Delta^+}^2}+\frac{1}{m_{\Delta^{++}}^2}\right)\,,
\label{AR2}\\
A_L(q^2) &=& -\,\frac{1}{\sqrt{2}\, G_F}\,
\frac{h_{le}^* h_{l\mu}}{6\pi^2}\,
\left( \frac{1}{12\,m_{\Delta^{+}}^2}\,
+ \,\frac{1}{m_{\Delta^{++}}^2}\, f\left(\frac{-q^2}{m_{\Delta^{++}}^2},\,
\frac{m_l^2}{m_{\Delta^{++}}^2}\right)
\right)
\label{AL2}\,,
\end{eqnarray}
%
$m_l$  being the mass of the charged lepton $l$, $l=e,\mu,\tau$.
 In the limit where the transition is dominated
by the exchange of a virtual doubly charged scalar  $\Delta^{++}$,
these expressions reduce to those obtained in
\cite{JBerna86,Raidal:1997hq};  to the best of our knowledge
the expression of $A_L(q^2)$ for the general case is a new result.
The term with the form factor $A_R$ in eq.~(\ref{LefftypeII})
generates the $\mu \to e \gamma$ decay amplitude.
It corresponds to the contribution of the one loop
diagrams with virtual neutrino and $\Delta^{+}$
\cite{STPZee82} and with virtual charged lepton and
$\Delta^{++}$ \cite{JBerna86} (see also \cite{BiPet87}).
The second term involving the form factor
$A_{L}$, together with $A_{R}$, generates the $\mu - e$ conversion amplitude.
The loop function $f(r,s_l)$ is well known \cite{Raidal:1997hq}:
\begin{eqnarray}
	f(r,s_l) &=& \frac{4s_l}{r}\,+\,\log(s_l)\,+
\,\left(1-\frac{2s_l}{r}\right)\,\sqrt{1+\frac{4s_l}{r}}\,
\log\frac{\sqrt{r}+\sqrt{r+4s_l}}{\sqrt{r}-\sqrt{r+4s_l}}\,.
\end{eqnarray}
%
Notice that in the limit in which the charged lepton
masses $m_l$ are much smaller than the doubly-charged
scalar mass $m_{\Delta^{++}}$, one has
$f(r,s_l)\simeq \log(r) = \log(m_\mu^2/m_{\Delta^{++}}^2)$.
 For   $m_{\Delta^{++}} = (100 - 1000)$ GeV,
this is an excellent approximation for
$f(r,s_e)$, but cannot be used for $f(r,s_\mu)$ and $f(r,s_\tau)$.
The ratios  $f(r,s_e)/f(r,s_\mu)$
and $f(r,s_e)/f(r,s_\tau)$ change relatively little when
$m_{\Delta^{++}}$  increases from 100 GeV  to 1000 GeV,
and at $m_{\Delta^{++}}= 100~(1000)$ GeV take the values:
$f(r,s_e)/f(r,s_\mu) \cong 1.2~(1.1)$ and
$f(r,s_e)/f(r,s_\tau) \cong 2.1~(1.7)$.
More generally,  $f(r,s_l)$, $l=e,\mu,\tau$,
are monotonically (slowly) decreasing functions
of  $m_{\Delta^{++}}$~\footnote{Note that we have  $f(r,s_l) < 0$, $l=e,\mu,\tau$.}:
for $m_{\Delta^{++}}= 100~(1000)$ GeV we have, e.g.,
$f(r,s_e) \cong -13.7~(-18.3)$.

\begin{figure}[t]
\begin{center}
\includegraphics[width=16cm,height=10cm]{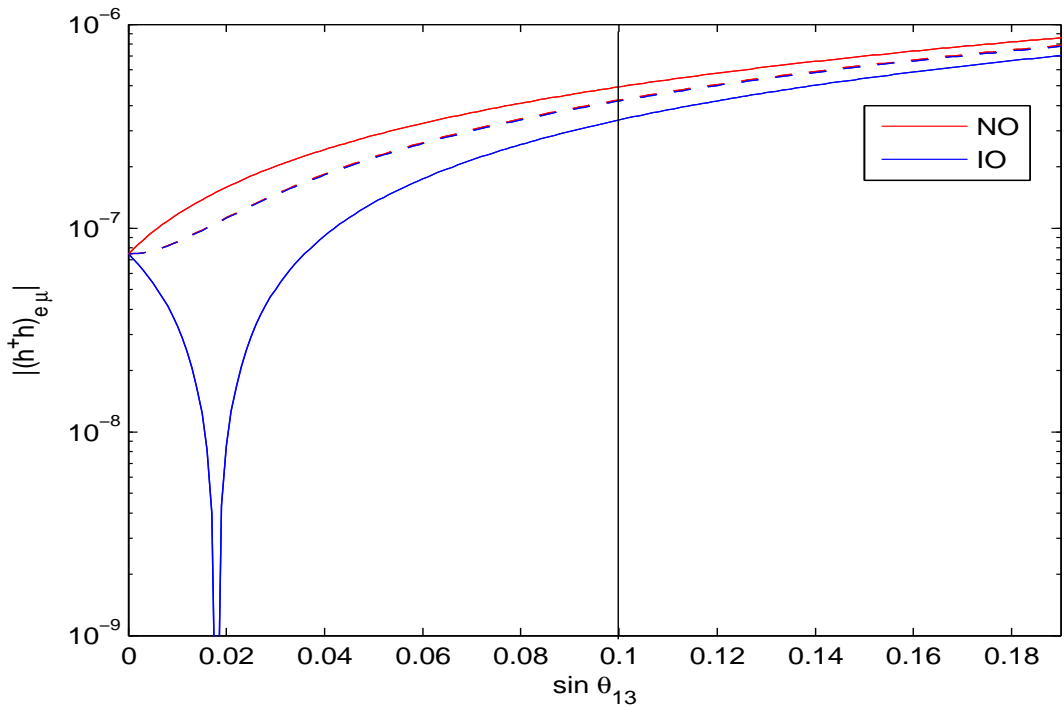}
\caption{The dependence of  $|(h^{\dagger}h)_{e\mu}|$ on
$\sin\theta_{13}$ for  $v_{\Delta} = 9.5$ eV
and $\delta = 0$ (solid lines) and $\delta = \pi/2$ (dashed line).
The other neutrino oscillation parameters are set to their best fit
values given in Table \ref{tab:tabdata-1106}.
The vertical line corresponds to the current $3\sigma$ allowed
minimal value of $\sin\theta_{13}$ (see eq.~(\ref{DBayth13})).
}
\label{fig:hhemuth13}
\end{center}
\end{figure}
%
%
\subsection{The  $\mu\to e \gamma$ Decay}\label{megII}
%
%
  Using eqs. (\ref{LefftypeII}) and (\ref{AR2}) we get for the
 $\mu\to e \gamma$ decay  branching ratio in the case
under discussion \cite{STPZee82,Akeroyd:2009nu}:
\begin{eqnarray}
{\rm BR}(\mu\to e \gamma) & \cong &
384\,\pi^2\,(4\pi\,\alpha_{\rm em}) \left|A_R\right|^2\;
=\; \frac{\alpha_{\rm em}}{192\,\pi}\,
\frac{\left|\left(h^{\dagger}h\right)_{e \mu}\right|^{2}}
{G_{F}^{2}}\,\left(
\frac{1}{m^2_{\Delta^{+}}} +
\frac{8}{m^2_{\Delta^{++}}}
\right)^{2}\,.
\label{muegII}
\end{eqnarray}
%
For $m_{\Delta^{+}} \cong  m_{\Delta^{++}} \equiv M_{\Delta}$,
the upper limit on  ${\rm BR}(\mu\to e \gamma)$
reported by the MEG experiment, eq.~(\ref{mutoegexp}),
implies the following upper bound on $|(h^{\dagger}h)_{e \mu}|$:
\begin{equation}
	\left|\left(h^{\dagger}h\right)_{e\mu}\right|\;<\;5.8\times 10^{-6}\,
\left(\frac{M_{\Delta}}{100\,{\rm GeV}}\right)^{2}\,.
\label{hhmueg}
\end{equation}
%
 One can use this upper bound, in particular, to obtain a lower bound
on the vacuum expectation value of $\Delta^0$, $v_{\Delta}$~\footnote{This was noticed also in \cite{Rodh}.}. Indeed,
from eq.~(\ref{hU}) it is not difficult to get:
\begin{equation}
\left |\left(h^{\dagger}h\right)_{e\mu}\right| = \frac{1}{4v^2_{\Delta}}\,
\left | U_{e2}\,U^{\dagger}_{2\mu}\, \Delta m^2_{21} +
 U_{e3}\,U^{\dagger}_{3\mu}\, \Delta m^2_{31} \right |\,,
\label{hhmue}
\end{equation}
%
where we have used the unitarity of $U$.
The above expression for $|(h^{\dagger}h)_{e\mu}|$ is
exact. It follows from eq.~(\ref{hhmue}) that the prediction
for $|(h^{\dagger}h)_{e\mu}|$, and thus for
${\rm BR}(\mu\to e \gamma)$, depends, in general,
on the Dirac CPV phase $\delta$ of the standard
parametrisation of the PMNS matrix $U$
(see eq.~(\ref{UPMNS})).
For the best fit values of $\sin^2\theta_{13} = 0.0236$
and of the other neutrino oscillation
parameters listed in Table \ref{tab:tabdata-1106},
the term $\propto\Delta m^2_{21}$
in eq.~(\ref{hhmue}) is approximately a factor of 10 smaller
than the term $\propto\Delta m^2_{31}$.
In this case ${\rm BR}(\mu\to e \gamma)$ exhibits a
relatively weak dependence
on the type of the neutrino mass spectrum
and on the Dirac phase $\delta$.
Neglecting the term $\propto\Delta m^2_{21}$,
we obtain from (\ref{hhmueg}) and (\ref{hhmue}):
\begin{equation}
v_{\Delta} > 2.1\times 10^2\,\left |
 s_{13}\,s_{23}\, \Delta m^2_{31} \right|^{\frac{1}{2}}\,
\left(\frac{100\,{\rm GeV}}{M_{\Delta}}\right)\,
\cong 3.0\,{\rm eV}\, \left(\frac{100\,{\rm GeV}}{M_{\Delta}}\right)\,.
\label{vDll}
\end{equation}
%
For the $3\sigma$ allowed ranges of values of
$\sin^22\theta_{13}$ given in eq.~(\ref{DBayth13})
and of the other neutrino oscillation
parameters quoted in Table \ref{tab:tabdata-1106},
the absolute lower bound on $v_{\Delta}$ corresponds
approximately to
$v_{\Delta} > 1.5~{\rm eV}~(100\,{\rm GeV})/M_{\Delta}$
and is reached in the case of
$\Delta m^2_{31} > 0$ ($\Delta m^2_{31} < 0$)
for $\delta = \pi~(0)$.

 We note further that if $\delta \cong \pi/2~(3\pi/2)$,
the term $\propto\Delta m^2_{21}$ in the expression
for  $|(h^{\dagger}h)_{e\mu}|$ (and thus for
${\rm BR}(\mu\to e \gamma)$) always plays a subdominant
role as long as the other neutrino oscillation
parameters lie in their currently allowed $3\sigma$
ranges. Therefore in this case the dependence of
${\rm BR}(\mu\to e \gamma)$ on the type of neutrino
mass spectrum is negligible.
The specific features of the predictions
for $|(h^{\dagger}h)_{e\mu}|$
discussed above are illustrated in Fig. \ref{fig:hhemuth13}.

 Exploiting the fact that  $v^2_{\Delta}|(h^{\dagger}h)_{e\mu}|$
is known with a rather good precision, we can write:
\begin{eqnarray}
{\rm BR}(\mu\to e \gamma) \cong
2.7\times 10^{-10}\, \left ( \frac{1\,{\rm eV}}{v_{\Delta}}\right)^4\,
\left (\frac{100\,{\rm GeV}}{M_{\Delta}}\right)^4\,,
\label{megIInum}
\end{eqnarray}
%
where we have used eq.~(\ref{muegII}) and the best
fit values of the neutrino oscillation parameters.
It follows from eq.~(\ref{megIInum}) that
for the values of $v_{\Delta}$ and
$M_{\Delta}$ (or $m_{\Delta^{+}}$ and/or
$m_{\Delta^{++}}$) of interest,
${\rm BR}(\mu\to e \gamma)$ can have a value
within the projected sensitivity of the
ongoing MEG experiment.

%
\subsection{The $\mu \rightarrow 3e $ decay}
%
%
  In the TeV scale type II see-saw scenario,
the $\mu\rightarrow 3e$ decay amplitude
is generated at the tree level by the diagram
with exchange of a virtual doubly-charged
Higgs boson  $\Delta^{++}$.
The branching ratio can be easily calculated
(see, e.g.,  \cite{JBerna86,BiPet87}):
\begin{equation}
{\rm BR}(\mu\rightarrow 3e) =
\frac{1}{G_F^2}\frac{|(h^\dagger)_{ee}(h)_{\mu e}|^2}{m^4_{\Delta^{++}}}\, =
\frac{1}{G_F^2 \,m^4_{\Delta^{++}}}\,
\frac{|m^*_{ee}\,m_{\mu e}|^2}{16\,v^4_{\Delta}}\,,
\label{IIBRmu3e}
\end{equation}
%
where we have used eq.~(\ref{mnuII}).
From the present limit ${\rm BR}(\mu\rightarrow 3e)<10^{-12}$,
one can obtain the following constraint on $|(h^+)_{ee}(h)_{\mu e}|$:
\begin{equation}
|(h^\dagger)_{ee}(h)_{\mu e}| < 1.2 \times 10^{-7}~
\left(\frac{m_{\Delta^{++}}}{100\,{\rm GeV}}\right)^{2}\,.
\end{equation}
%
\begin{figure}[t]
\begin{center}
\includegraphics[width=16cm,height=10cm]{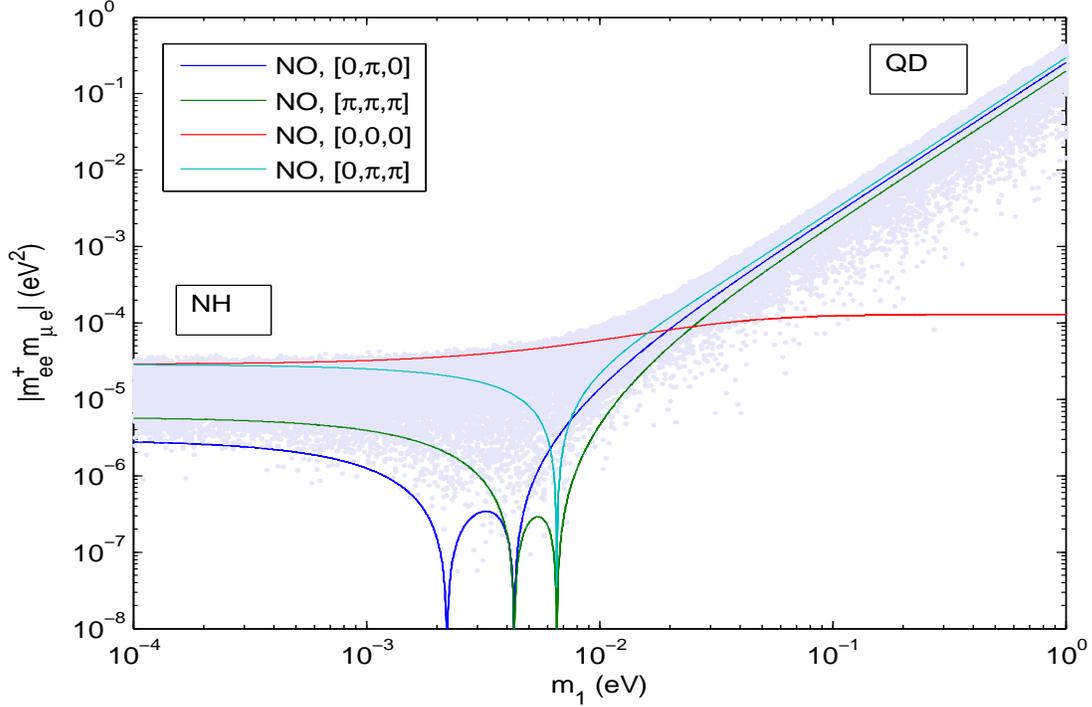}
\caption{The dependence of  $|m^*_{ee}m_{\mu e}|$ on the lightest
neutrino mass $m_1$ in the case of neutrino mass spectrum with
normal ordering ($\Delta m^2_{\rm A} > 0$),
for four sets of values of the Dirac
and the two Majorana
CPV phases, $[\delta,\alpha_{21},\alpha_{31}]$.
The depicted curves correspond to the best fit
values of $\sin\theta_{13}$ (eq.~(\ref{DBayth13}))
and of the other neutrino oscillation parameters
given in Table \ref{tab:tabdata-1106}.
The scattered points
are obtained by varying the neutrino oscillation parameters
within their corresponding $3\sigma$ intervals and giving
random values to the CPV Dirac and Majorana phases.}
\label{fig:MMNOII}
\end{center}
\end{figure}

\begin{figure}[t]
\begin{center}
\includegraphics[width=16cm,height=10cm]{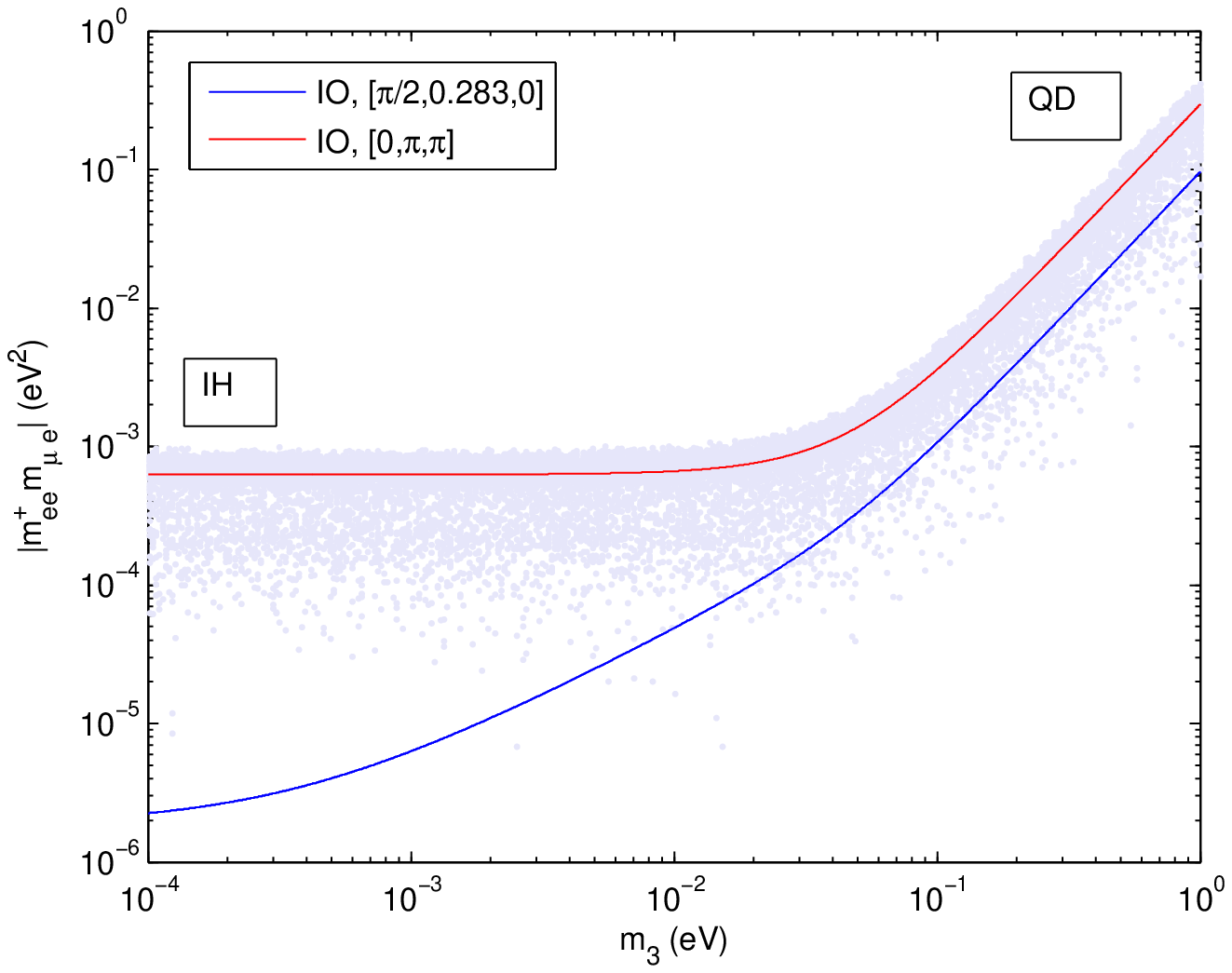}
\caption{
The same as in Fig.~\ref{fig:MMNOII}
in the case of a light neutrino mass spectrum with
inverted ordering ($\Delta m^2_{\rm A} <0$) (see text for details).
}
\label{fig:MMIOII}
\end{center}
\end{figure}
%

 In the model under discussion,
${\rm BR}(\mu\rightarrow 3e)$ depends
on the factor $|m^*_{ee}m_{\mu e}|$,
which involves the product of two
elements of the neutrino
Majorana mass matrix, on
the neutrino mass spectrum and
on the Majorana and Dirac
CPV phases in the PMNS
matrix $U$. For the values of
$m_{\Delta^{+}}$ and $m_{\Delta^{++}}$ in the range of
$\sim (100 - 1000)$ GeV and of  $v_{\Delta} \ll 1$ MeV
of interest, $m_{ee}$ practically coincides with
the effective Majorana mass
in neutrinoless double beta
($\betabeta$-) decay
(see, e.g., \cite{BiPet87,BPP1,PSugT08}),
$\mefff$:
\begin{equation}
|m_{ee}| =
\left | \sum_{j=1}^{3} m_j U^2_{e j}\right |
\cong \meff\,.
\label{hee}
\end{equation}
%
Depending on the type of neutrino mass spectrum,
the value of the lightest neutrino mass and
on the values of the CPV Majorana and Dirac phases in the
PMNS matrix, $|m_{ee}|$ can take any value between
0 and $m_0$, where $m_0 = m_{1} \cong m_2 \cong m_3$
is the value of the neutrino masses in the
case of quasi-degenerate (QD) spectrum,
$m_0\gtap 0.1$ eV (see, e.g., \cite{BPP1}).
It follows from the searches for the $\betabeta$-decay
that $|m_{ee}| \ltap m_0 \ltap 1$ eV,
while the cosmological constraints on the sum of the
neutrino masses imply $m_0\ltap 0.3$ eV
(see, e.g., \cite{PDG10}).
As is well known, the $\betabeta$-decay is
claimed to have been observed
in \cite{Klap04,KlapdorMPLA},
with the reported half-life
corresponding to \cite{KlapdorMPLA}
$|m_{ee}|= 0.32 \pm 0.03$~eV.
This claim will be tested in a new
generation of  $\betabeta$-decay experiments
which either are already taking data or
are in preparation at present
(see, e.g., \cite{PDG10,WRbb0nu11}).

 In the case of NH light neutrino mass spectrum with
$m_1 \ll 10^{-4}$ eV, $|m_{ee}|$ lies in the
interval  $3.6\times 10^{-4}~{\rm eV} \ltap |m_{ee}| \ltap
5.2\times 10^{-3}$ eV. This interval was obtained
by taking into account the
$3\sigma$ allowed ranges of values of
$\sin^2\theta_{13}$ (eq.~(\ref{DBayth13})),
$\sin^2\theta_{12}$, $\sin^2\theta_{23}$ and
$\Delta m^2_{\odot}$ and $\Delta m^2_{\rm A}$.
For the best fit values (b.f.v.) of the latter we get
\footnote{The numerical values quoted further in this
subsection are obtained for the indicated best fit
values of the neutrino oscillation parameters,
unless otherwise stated.}:
 $1.45\times 10^{-3}~{\rm eV} \ltap |m_{ee}| \ltap
3.75\times 10^{-3}$ eV.
The minimal and the maximal values correspond to the
combination of the CPV phases
$(\alpha_{21} -\alpha_{31} + 2\delta) = \pi~{\rm and}~0$,
respectively. However, for  $m_1 \gtap 10^{-4}$ eV,
one can have  $|m_{ee}| = 0$ for specific values of
$m_1$ if the CPV phases $\alpha_{21}$
and $\alpha_{31} - 2\delta$ possess the CP conserving values
$\alpha_{21} = \pi$ and $(\alpha_{31} - 2\delta) = 0,\pi$
(see, e.g., \cite{Pascoli:2007qh}):
for the $[\pi,0]$ combination this occurs at
$m_1 \cong 2.3\times 10^{-3}$ eV, while in the case
of the $[\pi,\pi]$ one we have $|m_{ee}| = 0$ at
$m_1 \cong 6.5\times 10^{-3}$ eV.
If the light neutrino mass spectrum is with inverted ordering
($\Delta m^2_{\rm A}\equiv \Delta m^2_{32} < 0$, $m_3 < m_1 < m_2$)
or of inverted hierarchical (IH) type ($m_3 \ll m_1 < m_2$),
we have~\cite{PPSNO02} $|m_{ee}| \gtap
\sqrt{|\Delta m^2_{\rm A}| + m^2_3}\cos2\theta_{12}
\gtap 1.27\times 10^{-2}$ eV, while in the case of
QD spectrum,
$|m_{ee}| \gtap m_0\cos2\theta_{12}\gtap  2.8\times 10^{-2}$ eV,
where we used the 3$\sigma$ minimal allowed values
of $|\Delta m^2_{\rm A}|$ and $\cos2\theta_{12}$.

 We consider next briefly the dependence of the
neutrino mass matrix element
 $|m_{\mu e}|$ on the type of the neutrino mass spectrum
and on the CPV Majorana and Dirac phases.
In the case of NH spectrum with $m_1 =0$,
the maximal value of $|m_{\mu e}|$ is
obtained for $\alpha_{31} - \alpha_{21} = \delta$,
$\delta = \pi$, and
reads: ${\rm max}(|m_{\mu e}|) \cong 8.1\times 10^{-3}$ eV.
We get  $|m_{\mu e}| = 0$
for  $\alpha_{21} = \pi$, $\delta = 0~(\pi)$ and  $\alpha_{31} = 0~(\pi)$.
As can be shown, for each of these two sets of values of the CPV
phases, the zero takes place at essentially the same value of
$m_1\cong 4.3\times 10^{-3}$ eV
(Fig. \ref{fig:MMNOII}).
If the neutrino mass spectrum is of the IH type with
negligible $m_3\cong 0$, the maximal value of
$|m_{\mu e}|$ corresponds to $\delta = 0$ and
$\alpha_{21} = \pi$ and is given by
${\rm max}(|m_{\mu e}|) \cong \sqrt {|\Delta m^2_{\rm A}|}\, c_{13}
(c_{23}\sin 2\theta_{12} + s_{23}s_{13}\cos2\theta_{12})$.
The element $|m_{\mu e}|$ is strongly suppressed,
i.e., we have  $|m_{\mu e}| \ll {\rm max}(|m_{\mu e}|)$,
for $\delta \cong \pi/2$ and a value of the
Majorana phase $\alpha_{21}$ which is determined
by the equation:
\begin{equation}
c_{23}\,c_{12}\,s_{12}\,\sin\alpha_{21} \cong
\left (c^2_{12} + s^2_{12}\cos\alpha_{21}\right)\,s_{23}\,s_{13}\,.
\label{hemu0IH}
\end{equation}
%
For the b.f.v. of the neutrino mixing angles
this equation is satisfied for $\alpha_{21} \cong 0.283$.

 The properties of $|m_{ee}|$ and $|m_{\mu e}|$ described above
allow us to understand most of the specific features of the
dependence of the quantity $|m^*_{ee}m_{\mu e}|$ of interest
on the the neutrino mass spectrum and the leptonic CPV phases.
For NH spectrum and negligible $m_1\cong 0$,
the maximum of the latter is obtained for
$\alpha_{31} - \alpha_{21} = \delta = 0$ and is given by:
\begin{equation}
{\rm max}(|m^*_{ee}m_{\mu e}|) =
\left |\left( m_2\,s^2_{12}\,c^2_{13} + m_3\, s^2_{13}\right)\,
c_{13}\left( m_2\,s_{12}(c_{12}\,c_{23} - s_{12}\,s_{23}\,s_{13}) +
m_3\,s_{23}\,s_{13}\right)\right|\,,
\label{maxmeemmueNH}
\end{equation}
%
with $m_2 =\sqrt{\Delta m^2_{\odot}}$ and
$m_3 =\sqrt{\Delta m^2_{\rm A}}$.
Using the b.f.v. of the
neutrino oscillation parameters we get
${\rm max}(|m^*_{ee}m_{\mu e}|) \cong 2.9\times 10^{-5}~{\rm eV^2}$
(see Fig. \ref{fig:MMNOII}).
This implies ${\rm BR}(\mu\rightarrow 3e) \ltap
6\times 10^{-9}(1~{\rm eV}/v_{\Delta})^4
(100~{\rm GeV}/m_{\Delta^{++}})^4$.
In the case of NH spectrum and non-negligible $m_1$
we have  $|m^*_{ee}m_{\mu e}| = 0$ for the values of
the CPV phases and $m_1$ discussed above,
for which either $|m_{ee}| = 0$ or $|m_{\mu e}| = 0$.
The scattered points in Fig. \ref{fig:MMNOII}
correspond to the possible values
the quantity $|m^*_{ee}m_{\mu e}|$ can assume when
varying the neutrino oscillation parameters
within their corresponding $3\sigma$ intervals and giving
random values to the CPV Dirac and Majorana phases
from the interval [0,2$\pi$].

  The maximum of $|m_{ee}m_{\mu e}|$ for the IH spectrum
with a negligible $m_3$ is reached for $\delta = 0$ and
$\alpha_{21} = \pi$, and reads:
\begin{equation}
{\rm max}(|m^*_{ee}m_{\mu e}|) \cong
\left |\Delta m^2_{\rm A}\right |\, c^3_{13}\,
\left (\frac{1}{2}\,c_{23}\,\sin 4\theta_{12} +
s_{23}\,s_{13}\,\cos^22\theta_{12}\right )\,.
\label{maxmeemmueIH}
\end{equation}
%
Numerically this gives
${\rm max}(|m_{ee}m_{\mu e}|) \cong 6.1\times 10^{-4}~{\rm eV^2}$
(Fig. \ref{fig:MMIOII}).
For  ${\rm BR}(\mu\rightarrow 3e)$ we thus obtain:
${\rm BR}(\mu\rightarrow 3e) \ltap
2.4\times 10^{-6}(1~{\rm eV}/v_{\Delta})^4
(100~{\rm GeV}/m_{\Delta^{++}})^4$.
One can have $|m_{ee}m_{\mu e}|\ll {\rm max}(|m_{ee}m_{\mu e}|)$
in the case of IH spectrum with $m_3=0$
for, e.g., $\delta \cong \pi/2$ and $\alpha_{21} \cong 0.283$,
for which  $|m_{\mu e}|$ has a minimum.
For the indicated values of the phases we find:
$|m_{ee}m_{\mu e}| \cong 1.2\times 10^{-6}~{\rm eV^2}$
(see Fig.~\ref{fig:MMIOII}).
Similarly to the case of a neutrino mass spectrum
with normal ordering discussed above,
we show in Fig.~\ref{fig:MMIOII}
the range of values the LFV term
$|m_{ee}m_{\mu e}|$ can assume
(scattered points).

 Finally, in the case of QD spectrum,
$|m_{ee}m_{\mu e}|$ will be relatively
strongly suppressed with respect to its possible
maximal value for this spectrum (i.e., we will have
$|m_{ee}m_{\mu e}| \ll {\rm max}(|m^*_{ee}m_{\mu e}|)$)
if, e.g., the Majorana and Dirac phases are zero,
thus conserving the CP symmetry:
$\alpha_{21} = \alpha_{31} = \delta = 0$.
Then one has: $|m_{ee}m_{\mu e}|\cong
|\Delta m^2_{\rm A}|s_{13}s_{23}c_{13}/2
\cong 1.2\times 10^{-4}~{\rm eV^2}$.
Note that this value is still larger than the
maximal value of $|m_{ee}m_{\mu e}|$
for the NH neutrino mass spectrum with a
negligible $m_1$ (see Fig. \ref{fig:MMNOII}).
The maximum of $|m_{ee}m_{\mu e}|$
takes place for another set of
CP conserving values of the Majorana and Dirac phases:
$\alpha_{21} = \alpha_{31} = \pi$ and
$\delta = 0$. At the maximum we have:
\begin{equation}
{\rm max}(|m^*_{ee}m_{\mu e}|) \cong
m^2_0\,\left ( c^3_{13}\,\cos2\theta_{12} - s^2_{13}\right )\,
c_{13}\,\left (c_{23}\,\sin 2\theta_{12} +
2\,c^2_{12} s_{23}\,s_{13}\right )\,,~~m_0\gtap 0.1~{\rm eV}\,.
\label{maxmeemmueQD}
\end{equation}
%
For the b.f.v. of the neutrino mixing angles we get
${\rm max}(|m^*_{ee}m_{\mu e}|)\cong 0.3\,m^2_0$.
For $m_0\ltap 0.3$ eV this implies
${\rm max}(|m^*_{ee}m_{\mu e}|)\ltap 2.7\times 10^{-2}~{\rm eV^2}$,
leading to an upper bound on
${\rm BR}(\mu\rightarrow 3e)$, which is by
a factor approximately of $4.1\times 10^3$
larger than in the case of IH spectrum.

 The features of  $|m_{ee}m_{\mu e}|$ discussed above are
illustrated in Figs. \ref{fig:MMNOII} and \ref{fig:MMIOII}.

 \vspace{0.4cm}
 It should be clear from the preceding discussion that
in the case of the type II see-saw model considered,
the value of the quantity
$|(h^\dagger)_{ee}(h)_{\mu e}|^2 \propto |m^*_{ee}m_{\mu e}|^2$,
and thus the prediction for ${\rm BR}(\mu\rightarrow 3e)$,
depends very strongly on the type of neutrino
mass spectrum. For a given spectrum,
it exhibits also a very strong dependence on the values
of the Majorana and Dirac CPV phases $\alpha_{21}$,
$\alpha_{31}$ and $\delta$, as well as on the value of the
lightest neutrino mass, ${\rm min}(m_j)$.  As a consequence,
the prediction for  ${\rm BR}(\mu\rightarrow 3e)$
for given $v_{\Delta}$ and $m_{\Delta^{++}}$ can vary by
a few to several orders of magnitude
when one varies the values of  ${\rm min}(m_j)$ and
of the CPV phases.
Nevertheless, for all possible types of neutrino
mass spectrum - NH, IH, QD, etc.,
there are relatively large regions of the parameter
space of the model where ${\rm BR}(\mu\rightarrow 3e)$
has a value within the sensitivity of the
planned experimental searches for the
$\mu\rightarrow 3e$ decay  \cite{muto3eNext}.
The region of interest for the NH spectrum is
considerably smaller than those for the IH and QD
spectra. In the case NO spectrum
($\Delta m^2_{\rm A} > 0$),  ${\rm BR}(\mu\rightarrow 3e)$
can be strongly suppressed for certain values of the
lightest neutrino mass $m_1$ from the
interval $\sim (2\times10^{-3} - 10^{-2})$ eV
(Fig. \ref{fig:MMNOII}). For the IO spectrum
($\Delta m^2_{\rm A} < 0$), a similar suppression
can take place for $m_3 \ll 10^{-2}$ eV (Fig. \ref{fig:MMIOII}).
In the cases when $|m^*_{ee}m_{\mu e}|^2$ is very
strongly suppressed, the one-loop corrections to the
$\mu\rightarrow 3e$ decay amplitude should be taken
into account since they might give a larger
contribution than that of the tree level
diagram we are considering.
The analysis of this case, however, is beyond
the scope of the present investigation.

%
\subsection{The $\mu - e$ Conversion in Nuclei}
%

 Consider next the $\mu-e$ conversion in a generic nucleus
$\mathcal{N}$. We parametrise the corresponding
conversion rate following the
effective field theory approach developed in
\cite{Kitano:2002mt}. Taking into account the
interaction Lagrangian (\ref{LefftypeII}),
we get in the type II see-saw scenario
\begin{eqnarray}
 {\rm CR}(\mu\, \mathcal{N}\to e\, \mathcal{N})
&\cong &
(4\pi\alpha_{\rm em})^2\,\frac{2 \,G_{F}^{2}}{\Gamma_{\rm capt}}\,
\left| A_{R}\,\frac{D}{\sqrt{4\pi\,\alpha_{\rm em}}}\, +
\,(2\,q_{u}+q_{d})\,A_L\,V^{(p)}\right|^{2}\,.
\label{CRtypeII}
\end{eqnarray}
%
The parameters  $D$ and $V^{(p)}$ represent overlap
integrals of the muon and electron wave functions and are related
to the effective dipole and vector type operators in
the interaction Lagrangian, respectively (see, $e.g.$ \cite{Kitano:2002mt}).

In the case of a light nucleus, $i.e.$ for $Z\lesssim 30$, one has with
a good approximation $D\simeq 8\,\sqrt{4\pi\alpha_{\rm em}}\,V^{(p)}$,
with the vector type overlap integral of the proton,
$V^{(p)}$, given by \cite{Kitano:2002mt}:
\begin{equation}
	V^{(p)}\;\simeq\; \frac{1}{4\pi}\,m_{\mu}^{5/2}\,
\alpha_{\rm em}^{3/2}\,Z_{eff}^{2}\,Z^{1/2}\,F(-m_{\mu}^{2})\,,
\label{Vp}
\end{equation}
%
where $F(q^2)$ is the form factor of the nucleus.
The parameters $D\,m_{\mu}^{-5/2}$, 	
$V^{(p)}\,m_{\mu}^{-5/2}$ and $\Gamma_{\rm capt}$
for $_{22}^{48}\text{Ti}$, $_{13}^{27}\text{Al}$ and
$_{79 }^{197}\text{Au}$ are given in Table \ref{NuclParam}.
\begin{table}[t]
\begin{center}
\begin{tabular}{|ccccc|}
\hline \hline
\rule[0.15in]{0cm}{0cm}{\tt $\mathcal{N}$}& $D\;m_{\mu}^{-5/2}$ & $V^{(p)}\;m_{\mu}^{-5/2}$ & $V^{(n)}\;m_{\mu}^{-5/2}$ & $\Gamma_{\rm capt}~(10^{6}\,\text{s}^{-1})$ \\
\hline
\rule[0.25in]{0cm}{0cm}$^{48}_{22}$Ti & 0.0864 & 0.0396 & 0.0468 & 2.590\\
\rule[0.25in]{0cm}{0cm}$^{27}_{13}$Al & 0.0362  & 0.0161 & 0.0173 & 0.7054\\
\rule[0.25in]{0cm}{0cm}$^{197}_{79}$Au & 0.189 & 0.0974 & 0.146  & 13.07 
\\\hline\hline
\end{tabular}
\caption{Nuclear parameters related to $\mu-e$ conversion in
$_{22}^{48}\text{Ti}$, $_{13}^{27}\text{Al}$ and
$_{79 }^{197}\text{Au}$. The numerical values of
the overlap integrals $D$, $V^{(p)}$ and $V^{(n)}$
are taken from \cite{Kitano:2002mt}.}
\label{NuclParam}
\end{center}
\end{table}%
%
Using the result for $D$ quoted above,
eqs. (\ref{AR2}), (\ref{AL2}) and (\ref{Vp}),
the conversion rate (\ref{CRtypeII}) can be written as
\begin{eqnarray}
	 {\rm CR}(\mu\, \mathcal{N}\to e\, \mathcal{N}) &\cong&
\frac{\alpha^5_{\rm em}}{36\,\pi^4}\,
\frac{m_\mu^5}{\Gamma_{\rm capt}}\,
Z_{eff}^4\,Z\,F^2(-m_{\mu}^{2})\,
\left| \left(h^{\dagger}h\right)_{e\mu }\,
\left [ \frac{5}{24\,m^2_{\Delta^{+}}}
+ \frac{1}{m^2_{\Delta^{++}}}\right ] \right.\nonumber \\
  & + & \left. \frac{1}{m^2_{\Delta^{++}}}\,
\sum_{l=e,\mu,\tau} h^{\dagger}_{e l}\,f(r,s_l)\, h_{l\mu}
\right |^2\,.
\label{CRtypeII2}
\end{eqnarray}
%
In contrast to previous studies, which assume that the $\mu-e$ conversion is dominated by the $\Delta^{++}$ exchange~\cite{Raidal:1997hq}, we will consider in this work a scenario where both scalars contribute to the transition amplitude. Thus, assuming $m_{\Delta^{+}}\cong m_{\Delta^{++}} \equiv M_{\Delta}$,
we have  ${\rm CR}(\mu\, \mathcal{N}\to e\, \mathcal{N})\propto
|C^{(II)}_{\mu e}|^2$, where
\begin{equation}
C^{(II)}_{\mu e}\equiv \frac{1}{4v^2_{\Delta}}
\left[\frac{29}{24}\,\left(m^{\dagger}\,m\right)_{e\mu }
+\,\sum_{l=e,\mu,\tau} m^{\dagger}_{e l}\,f(r,s_l)\, m_{l\mu}\,
\right ]\,,
\label{CmueII}
\end{equation}
%
and we have used eq.~(\ref{mnuII}).
The upper limit on the  $\mu - e$ conversion rate in Ti,
eq.~(\ref{mu2eTi}), leads to the following upper
limit on  $|C^{(II)}_{\mu e}|$:
\begin{equation}
|C^{(II)}_{\mu e}| < 1.24 \times 10^{-4}~
\left(\frac{M_{\Delta}}{100\,{\rm GeV}}\right)^{2}\,.
\label{CmueIIlimit}
\end{equation}
%
In obtaining it we have used the values of
$\Gamma_{\rm capt}$, $Z_{eff}$, $Z$ and
$F(-m_{\mu}^{2})$ for Ti given in subsection 2.2.
An experiment sensitive to
${\rm CR}(\mu\, {\rm Ti} \to e\, {\rm Ti})\approx 10^{-18}$~\cite{PRIME}
will be able to probe values of
$|C^{(II)}_{\mu e}| \gtap 5.8 \times 10^{-8}~
(M_{\Delta}/(100\,{\rm GeV}))^{2}$.

\begin{figure}[t]
\begin{center}
\includegraphics[width=16cm,height=10cm]{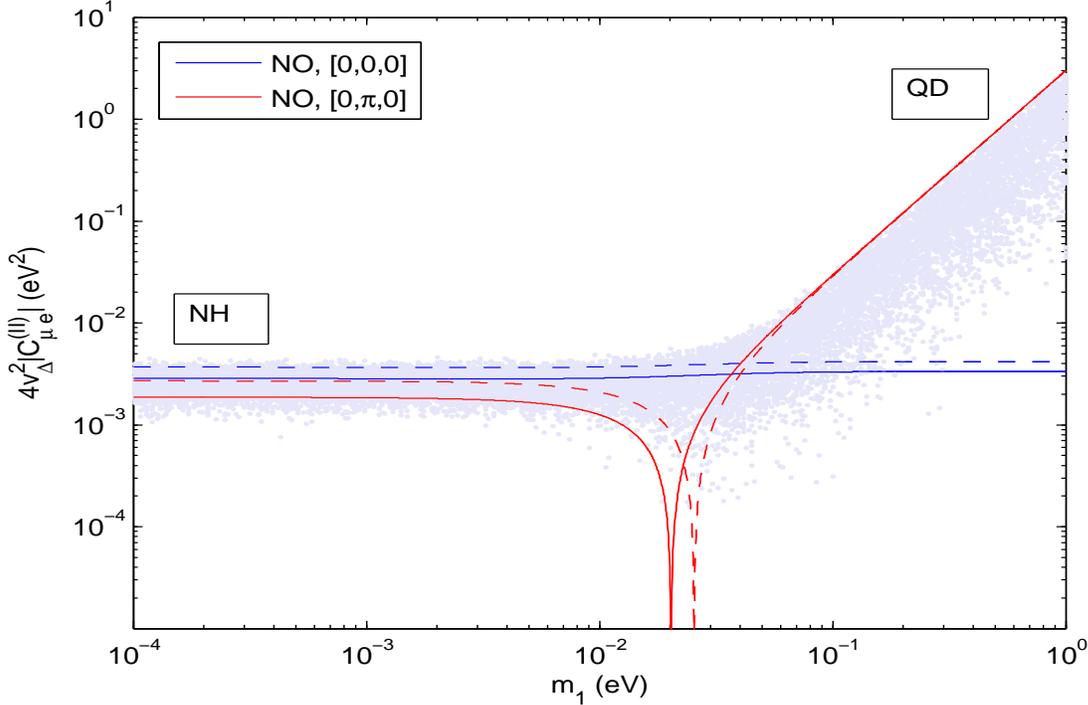}
\caption{
The dependence of $4v^2_{\Delta}|C^{(II)}_{\mu e}|$
(given in eV$^2$)  on the lightest
neutrino mass $m_1$ in the case of
neutrino mass spectrum with
normal ordering ($\Delta m^2_{\rm A} > 0$),
for two sets of values of the Dirac
and the two Majorana
CPV phases, $[\delta,\alpha_{21},\alpha_{31}]$
and $M_{\Delta}=200\;(1000)$ GeV, plain (dashed) curves.
The figure is obtained for the best fit
values of $\sin\theta_{13}$ (eq.~(\ref{DBayth13}))
and of the other neutrino oscillation parameters
given in Table \ref{tab:tabdata-1106} (see text for details).
}
\label{fig:NOCII}
\end{center}
\end{figure}

\begin{figure}[t]
\begin{center}
\includegraphics[width=16cm,height=10cm]{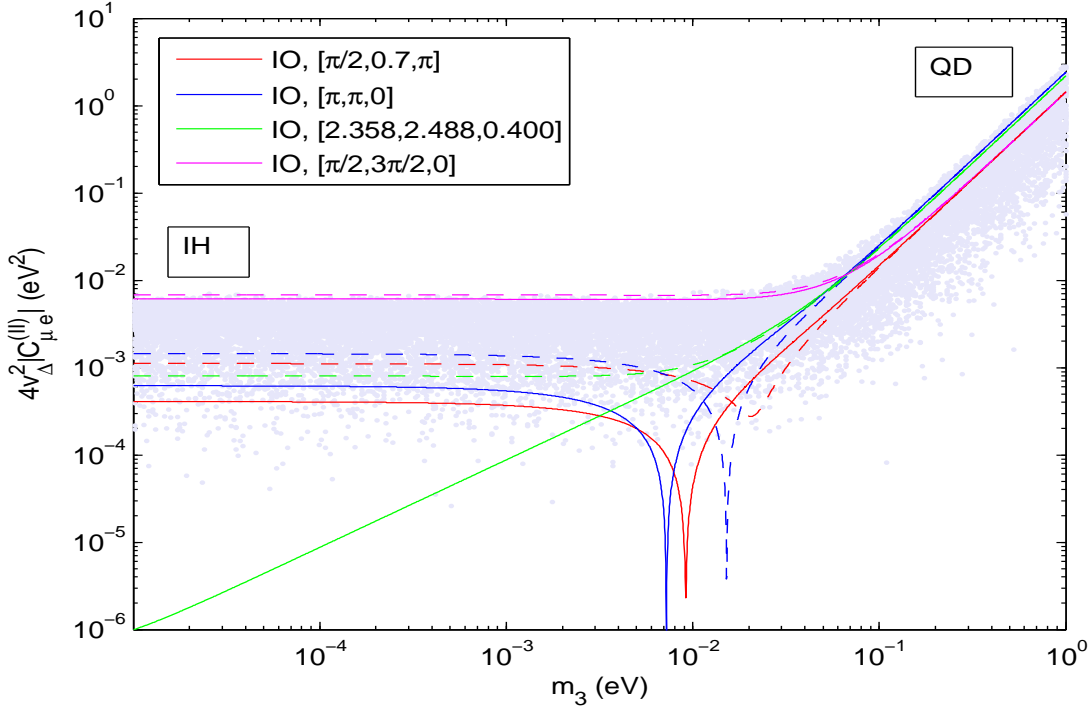}
\caption{
The same as in Fig.~\ref{fig:NOCII}
in the case of a light
neutrino mass spectrum with inverted ordering.
}
\label{fig:IOCII}
\end{center}
\end{figure}

    The  $\mu - e$ conversion rate in a given nucleus
depends through the quantity $C^{(II)}_{\mu e}$,
on the type of neutrino mass spectrum and the
Majorana and Dirac CPV phases in the PMNS matrix.
Using the b.f.v. of the the neutrino
oscillation parameters and performing a scan over the
values of the CPV phases and the lightest neutrino mass,
which in the cases of NO ($\Delta m^2_{\rm A} > 0$)
and IO ($\Delta m^2_{\rm A} < 0$) spectra
was varied in the intervals
$(10^{-4} - 1)$ eV and $(10^{-5} - 1)$ eV,
respectively, we have identified
the possible ranges of values of
$4v^2_{\Delta}|C^{(II)}_{\mu e}|$. The latter are
shown in Figs. (\ref{fig:NOCII}) and (\ref{fig:IOCII}).
%

For $M_{\Delta}=200~(1000)$ GeV and NH spectrum with negligible $m_1$
($m_1 \ll 10^{-3}$ eV),
the maximal value of $4v^2_{\Delta}|C^{(II)}_{\mu e}|$
occurs for $[\delta,\alpha_{21},\alpha_{31}] = [0,0,0]$ and at
the maximum we have
$4v^2_{\Delta}|C^{(II)}_{\mu e}| \cong 2.9~(3.8)\times 10^{-3}~{\rm eV^2}$.
For values of the CPV phases
$[\delta,\alpha_{21},\alpha_{31}] = [0,\pi,0]$
and $M_{\Delta}=200$ GeV,
$4v^2_{\Delta}|C^{(II)}_{\mu e}|$ goes
through zero at $m_1 \cong  2\times 10^{-2}~{\rm eV}$
(Fig. \ref{fig:NOCII}). In the case of a larger
charged scalar mass, $i.e.$ $M_{\Delta}=1000$ GeV,
such cancellation occurs at a different
value of the lightest neutrino mass, mainly $m_{1}=0.025$ eV.

 The maximum of $4v^2_{\Delta}|C^{(II)}_{\mu e}|$
in the case of IH spectrum
with negligible $m_3$,
occurs for maximal CPV phases:
$[\delta,\alpha_{21},\alpha_{31}] = [\pi/2,3\pi/2,0]$.
At the maximum in this case
one has $4v^2_{\Delta}|C^{(II)}_{\mu e}|
\cong 6~(7)\times 10^{-3}~{\rm eV^2}$ for $M_{\Delta}=200~(1000)$ GeV.
As Fig. \ref{fig:IOCII} shows,
for other sets of values of the CPV phases,
$4v^2_{\Delta}|C^{(II)}_{\mu e}|$ can be much smaller.
Taking again CP conserving phases, $e.g.$ $[\pi,\pi,0]$,
one can get a strong suppression of the branching ratio
for $m_{3}=7.2~(15)\times 10^{-3}$ eV and
$M_{\Delta}=200~(1000)$ GeV.
Allowing $\sin\theta_{13}$ to take values other
than the best fit one, we find that
$4v^2_{\Delta}|C^{(II)}_{\mu e}|$ can even go through zero
at, e.g., $[\delta,\alpha_{21},\alpha_{31}] = [\pi,\pi,\pi/2]$
for $\sin\theta_{13}\cong 0.137$, which lies within
the $2\sigma$ allowed region.
In Fig.~\ref{fig:IOCII} we report other examples in which
the CPV phases in the PMNS matrix
take different sets of CP violating values and
the quantity $4v^2_{\Delta}|C^{(II)}_{\mu e}|$
(and the conversion rate) can vary by several
orders of magnitude for specific
values of the lightest neutrino mass $m_{3}$
and the see-saw mass scale $M_{\Delta}$.

  If the neutrino mass spectrum  is quasi-degenerate,
$m_{1,2,3}\cong m_0 \gtap 0.1$ eV, we have for
$m_0 \ltap 0.3$ eV:
$2.8\times 10^{-3}~{\rm eV^2}\ltap 4v^2_{\Delta}|C^{(II)}_{\mu e}|
\ltap 0.4~{\rm eV^2}$. The minimal value corresponds to
$\Delta m^2_{\rm A} > 0$ (NO spectrum) and
$[\delta,\alpha_{21},\alpha_{31}] = [\pi,0,0]$;
for $e.g.$ $[\delta,\alpha_{21},\alpha_{31}] =
[0,0,0]$ and $M_{\Delta}=200$ GeV
we get in the QD region
$4v^2_{\Delta}|C^{(II)}_{\mu e}| \cong 3.3\times 10^{-3}~{\rm eV^2}$
(Fig. \ref{fig:NOCII}).

Finally, the scattered points
in Figs.~  \ref{fig:NOCII} and \ref{fig:IOCII}
are obtained by varying
all the neutrino oscillation parameters within the
corresponding $3\sigma$ intervals and allowing
for arbitrary values of
the Dirac and Majorana phases
in the interval [0,2$\pi$].

We remark that the previous estimates,
as well as Figs~\ref{fig:NOCII} and \ref{fig:IOCII},
were realized under the assumption
that the singly- and doubly-charged scalars have masses of
the same order, $i.e.$ $m_{\Delta^{+}}\cong m_{\Delta^{++}} \equiv M_{\Delta}$.
The case in which the dominant contribution
to the conversion amplitude is provided by the exchange of $\Delta^{++}$,
$i.e.$ for $m_{\Delta^{+}}\gg m_{\Delta^{++}}\gtrsim100$ GeV,
shows similar features: the upper limits of the conversion
ratio in the cases of NO and IO spectra are unchanged and a strong
suppression can occur for specific values of the CPV phases
and ${\rm min}(m_j)$.
Taking, instead, the opposite limit $m_{\Delta^{++}}\gg m_{\Delta^{+}}$,
with $m_{\Delta^{+}}=(100-1000)$ GeV, the dominant contribution to the
$\mu-e$ conversion amplitude is given by the exchange
of the singly-charged scalar, therefore we have:
$|C^{(II)}_{\mu e}| \propto |(h^{\dagger}h)_{e\mu}| $.
As it was pointed out in subsection~\ref{megII},
$|(h^{\dagger}h)_{e\mu}|$ shows a relative weak
dependance on the type of neutrino
mass spectrum and on the CPV phases in the PMNS matrix.
Moreover, no suppression of the conversion amplitude occurs
if $\sin(\theta_{13})$ is taken within
the current 3$\sigma$ experimental bound
(see Fig.~\ref{fig:hhemuth13}). In this case, from the
best experimental upper bound on the conversion rate in
Ti, ${\rm CR}(\mu\,\text{Ti}\to e\,\text{Ti})< 4.3\times 10^{-12}$,
we get the constraint:
\begin{equation}
	|(h^{\dagger}h)_{e\mu}| \;<\;6\times 10^{-4}\,\left( \frac{m_{\Delta^{+}}}{100\,\text{GeV}}\right)^{2}\,,
\end{equation}
%
which provides a weaker bound with respect to
that obtained from the $\mu\to e \gamma$ decay  (see eq.~(\ref{hhmueg})).
A $\mu-e$ conversion experiment sensitive to
$i.e.$  ${\rm CR}(\mu\,\text{Ti}\to e\,\text{Ti})\approx 10^{-18}$,
can probe values of  $|(h^{\dagger}h)_{e\mu}|$
which are by a factor $2\times 10^{3}$ smaller
and could set the limit:
\begin{equation}
|(h^{\dagger}h)_{e\mu}| \;<\;3\times 10^{-7}\,
\left( \frac{m_{\Delta^{+}}}{100\,\text{GeV}}\right)^{2}\,.
\end{equation}


%
\section{TeV Scale Type III See-Saw Model}\label{TypeIIISec}

  We turn in this section to the study of lepton
flavour violating processes in type III see-saw \cite{Foot:1988aq}
extensions of the SM. In the scenarios under discussion, the SM
particle content is enlarged by adding
$SU(2)_{L}$-triplets of fermions,
$\mbox{\boldmath $F$}_{jR}\equiv
\left(F_{jR}^{1},F_{jR}^{2},F_{jR}^{3}\right)$, $j\geq 2$,
possessing zero weak hypercharge and a
mass $M_{k}$ at the electroweak scale: $M_{k}\approx(100-1000)$ GeV.
The corresponding interaction and mass terms in the
see-saw Lagrangian read:
\begin{eqnarray}
\label{typeIIILag0}
\mathcal{L}^{\rm III}_{\rm seesaw} &=&
-\lambda_{\ell j} \,\overline{\psi}_{\ell L}\,
\mbox{\boldmath $\tau$}\,\widetilde{H} \cdot \,
\mbox{\boldmath $F$}_{jR}\,
- \,\frac{1}{2}\,\left(M_{R}\right)_{ij}\,
\overline{\mbox{\boldmath$F$}_{iL}^{C}}\cdot\mbox{\boldmath $F$}_{jR}\,
+ \; {\rm h.c.}\,,
\label{typeIIILag}
\end{eqnarray}
%
where $\mbox{\boldmath $\tau$}\equiv(\tau^{1},\tau^{2},\tau^{3})$,
$\tau^{a}$ being the usual $SU(2)_{L}$ generators
in the fundamental representation.

  It is convenient in the following discussion to work with the
charge eigenstates
$F_{jR}^{\pm}\equiv(F_{jR}^{1}\mp iF_{jR}^{2})$ and
$F_{jR}^{0}\equiv F_{jR}^{3}$.
Then, the physical states in the above
Lagrangian correspond to electrically
charged Dirac and neutral Majorana fermions,
which are denoted as $E_{j}$ and $N_{j}$, respectively:~\footnote{In the following we will
denote as $E_{j}$ and $N_{j}$ the mass eigenstates
obtained from the diagonalization of the full charged
and neutral lepton mass matrices.}
\begin{eqnarray}
&&E_{j}\;\equiv\; F_{jR}^{-}\,+\,F_{jL}^{+C}\,\quad\,
N_{j}\;\equiv\; F_{jL}^{0C}\,+\,F_{jR}^{0}\,.
\end{eqnarray}
%
In the basis in which the charged lepton mass matrix is diagonal,
the CC and NC weak interaction Lagrangian of
the light Majorana neutrino mass eigenstates $\chi_{j}$ read:
\begin{eqnarray}
\label{nuCCIII}
\mathcal{L}_{CC}^\nu
&=& -\,\frac{g}{\sqrt{2}}\,
\bar{\ell}\,\gamma_{\alpha}\,
\left( (1-\eta)U \right)_{\ell i}\,\chi_{i L}\,W^{\alpha}\;
+\; {\rm h.c.}\,,\\
\label{NCCIII}
\mathcal{L}_{NC}^\nu &=&  -\,\frac{g}{2 c_{w}}\,
\overline{\chi_{i L}}\,\gamma_{\alpha}\,
\left (U^\dagger(1+2\,\eta)U\right)_{ij}\,\chi_{j L}\,
Z^{\alpha}\,.
\label{NNCIII}
\end{eqnarray}
%
Similarly to the type I see-saw scenario discussed earlier,
the heavy Majorana mass eigenstates $N_{k}$
might acquire a sizable coupling to the weak gauge bosons
through the mixing with the light Majorana neutrinos:
\begin{eqnarray}
\label{NCCIII}
\mathcal{L}_{CC}^{N}
&=&
\,\frac{g}{2\sqrt{2}}\,
\bar{\ell}\,\gamma_{\alpha}\,(RV)_{\ell k}(1 - \gamma_5)\,N_{k}\,W^{\alpha}\;
+\; {\rm h.c.}\,,\\
\label{NNCIII}
\mathcal{L}_{NC}^{N} &=&
-\,\frac{g}{4 c_{w}}\,\overline{\nu}_{\ell}\,\gamma_{\alpha}\,(RV)_{\ell j}(1 - \gamma_5)\,N_{j}\,Z^{\alpha}\;+\;{\rm h.c.}\,.
\end{eqnarray}
%
In the expressions given above, the non-unitary
part of the neutrino mixing matrix, $i.e.$ the matrix  $\eta$,
and the matrix $R$ are defined as in the type I see-saw scenario
discussed in Section~\ref{TypeISec} (see eq.~(\ref{eta})), while $V$ in this case
diagonalizes the symmetric mass matrix $M_{R}$ in eq.~(\ref{typeIIILag}):
$M_{R}\cong V^{*}\diag(M_{1},M_{2},\ldots)V^{\dagger}$.

  The neutrino Yukawa couplings $\lambda_{\ell j}$ can
be partially constrained by low-energy neutrino oscillation data and
electroweak  precision observable
(see, $e.g.$ \cite{Abada:2007ux,Abada:2008ea}). Notice that, unlike
the type I see-saw extension of the Standard Model,
now we have flavour changing neutral currents~(FCNCs) in the charged lepton sector. The latter are
described by the interaction Lagrangian:
\begin{eqnarray}
\label{ellNCIII}
\mathcal{L}_{NC}^{\ell} &=& \,\frac{g}{2 c_{w}}\,
\left(\,\overline{\ell}_{L}\,\gamma_{\alpha}
\left(\mathbf{1}-4\,\eta\right)_{\ell \ell^{\prime}}\,\ell^{\prime}_{L}\,
-2\,s_{w}^{2}\,\overline{\ell}\,\gamma_{\alpha}\,\ell\,\right)\,Z^{\alpha}\,.
\end{eqnarray}
Finally,~\footnote{Flavour changing couplings between the charged leptons and
the SM Higgs boson $H$ arise as well in the TeV-scale type III see-saw
scenarios \cite{Abada:2008ea} which enter at one-loop
in the lepton flavour violating processes (see next subsection).}
the interactions of the new heavy charged leptons, $E_{j}$,
with the weak gauge bosons at leading order in
the mixing angle between the heavy and the light mass
eigenstates read:
\begin{eqnarray}
\mathcal{L}_{CC}^{E}
&=&\,-\,g\,\overline{E}_{j}\,\gamma_{\alpha}\,N_{j}\,W^{\alpha}\;
+\;g\,\overline{E}_{j}\,\gamma_{\alpha}\,(RV)_{\ell j}\,\nu_{\ell R}^{C}\,
W^{\alpha}\;+\; {\rm h.c.}\,,\\
\label{ENCIII}
\mathcal{L}_{NC}^{E} &=&
g\, c_{w}\,\overline{E}_{j}\,\gamma_{\alpha}\, E_{j}\,Z^{\alpha}\,
-\,\frac{g}{2\sqrt{2} c_{w}}\,
\left(\overline{\ell}\,\gamma_{\alpha}\,(RV)_{\ell j}(1-\gamma_{5})\,E_{j}\,
Z^{\alpha}\;+\;{\rm h.c.}\right)\,.
\end{eqnarray}

%
\subsection{The $\mu\to e \gamma$ Decay}
%
%
Charged lepton radiative decays receive additional
contributions with respect to the scenario with
singlet RH neutrinos, due to
the presence of new lepton flavour violating
interactions in the low energy effective Lagrangian
(see eqs.~(\ref{ellNCIII}) and (\ref{ENCIII})).
Following the computation reported in $\cite{Abada:2008ea}$,
we have for the $\mu\to e \gamma$ decay branching ratio
in the present scenario:
\begin{eqnarray}
{\rm BR}(\mu\to e \gamma)
&=&
\frac{3\alpha_{\rm em}}{32\pi}\,|T|^{2}\,,
\label{Bmutoegtype3}
\end{eqnarray}
%
where the amplitude $T$ is given by
\begin{equation}
T\;\cong\; -2\left(\frac{13}{3}+\mathcal{C}\right)\eta_{\mu e}\,
+\,\sum\limits_{k}\,(RV)_{ek}(RV)_{\mu k}^{*}\,
\left[ \,A(x_{k})+B(y_{k})+C(z_{k})\,\right]\,,
\end{equation}
%
with  $x_{k}=(M_{k}/M_{W})^{2}$, $y_{k}=(M_{k}/M_{Z})^{2}$,
$z_{k}=(M_{k}/M_{H})^{2}$ and
$\mathcal{C}\simeq -6.56$.
The loop functions $A(x_{k})$, $B(y_{k})$ and $C(z_{k})$
read \cite{Abada:2008ea}:
\begin{eqnarray}
A(x) &=& \frac{-30\,+\,153\, x\,-\,198\, x^{2}\,
+\,75\, x^{3}\,+\,18\,(4-3\,x)\,x^{2}\,\log x}{3(x-1)^{4}}\,,\\
B(y) &=& \frac{33\,-\,18\, y\,-\,45\, y^{2}\,+\,30\, y^{3}\,
+\,18\,(4-3\,y)\,y\,\log y}{3(y-1)^{4}}\,,\\
C(z) &=& \frac{-7\,+\,12\, z\,+\,3\, z^{2}\,
-\,8\, z^{3}\,+\,6\,(3\,z-2)\,z\,\log z}{3(z-1)^{4}}\,.
\end{eqnarray}
%
In the simple scenario of degenerate fermion triplets with an
overall mass scale $\overline{M}$ we obtain
taking $M_{H}=125$ GeV:
\begin{eqnarray}
	T/\eta_{\mu e}\;\cong\; \,11.6~(5.2)\,,
\quad{\rm for}\quad\overline{M}=100~(1000)~{\rm GeV}\,.
\end{eqnarray}
%
For $\overline{M}=100~(1000)$ GeV,
the current best upper limit on the $\mu\to e \gamma$ decay
branching ratio obtained in the MEG experiment,
eq.~(\ref{mutoegexp}),
implies the bound:
\begin{equation}
	|\eta_{\mu e}|\;<\;9~(20)\times 10^{-6}\,,
\quad{\rm for}\quad\overline{M}=100~(1000)~{\rm GeV}\,.
\label{MEGIII}
\end{equation}
%
If  no positive signal will be observed by  the MEG experiment, that is if it results that
${\rm BR}(\mu\to e \gamma)<10^{-13}$, the following upper limit on the non-unitarity lepton flavour violating coupling $|\eta_{\mu e}|$ can be set:
\begin{equation}
	|\eta_{\mu e}|\;<\;2~(4)\times 10^{-6}\,,
\quad{\rm for}\quad\overline{M}=100~(1000)~{\rm GeV}\,.
\label{MEGIIIbest}
\end{equation}

%
\subsection{The $\mu\to 3\,e$ and  $\mu-e$ Conversion in Nuclei}
%
%
The effective  $\mu-e-Z$ effective coupling in
the Lagrangian (\ref{ellNCIII})  provides the
dominant contribution (at tree-level) to the $\mu \to 3e$
decay rate and  the $\mu-e$ conversion rate in a nucleus.
In the case of the first process we have
(see, $e.g.$, \cite{Abada:2007ux}):
\begin{eqnarray}
	{\rm BR}(\mu \to 3e)
	&\simeq& 16\,|\eta_{\mu e}|^{2}\,
\left(3 \sin^{4}\theta_{W}\,-\,2\sin^{2}\theta_{W}\,+\,\frac{1}{2} \right)\,.
\end{eqnarray}
%
Taking into account the experimental upper limit
reported in (\ref{muto3eexp}), we get the following
upper limit on the $\mu-e$ effective coupling:
\begin{equation}
	|\eta_{\mu e}|\;<\; 5.6\times 10^{-7}\,.
\end{equation}
%
which is a stronger constraint with respect to
the one derived from the non-observation of
the $\mu\to e \gamma$ decay
(see eqs. (\ref{MEGIIIbest}) and (\ref{MEGIII})),
mediated (at one-loop) by an effective dipole operator.

More stringent constraints on the effective $\mu-e-Z$
coupling can be obtained using the data from the
$\mu-e$ conversion experiments.
Indeed, according to the general parametrisation given in
\cite{Kitano:2002mt} (see also \cite{Bernabeu:1993ta, Abada:2008ea}),
we have for the $\mu-e$ conversion ratio in a nucleus $\mathcal{N}$
with $N$ neutrons and $Z$ protons:
\begin{equation}
\label{CRtypeIII}
	 {\rm CR}(\mu\, \mathcal{N}\to e\, \mathcal{N})\;\cong\;
\frac{2 \,G_{F}^{2}}{\Gamma_{\rm capt}}\,|\mathcal{C}_{\mu e}|^{2}\,
\left|(2\,g_{LV(u)}+g_{LV(d)})\,V^{(p)}+(g_{LV(u)}+2\,g_{LV(d)})\,V^{(n)}
\right|^{2}\,,
\end{equation}
%
where in this case
\footnote{The expression for $V^{(n)}$ is valid under
the assumption of equal proton and neutron number
densities in the given nucleus \cite{Kitano:2002mt}.
The numerical value of the nuclear form factors for  $_{22}^{48}\text{Ti}$, $_{13}^{27}\text{Al}$ and
$_{79 }^{197}\text{Au}$ is reported in Table~\ref{NuclParam}. }
\begin{eqnarray}
	 \mathcal{C}_{\mu e}&\;\equiv&\;4\,\eta_{\mu e}\,,
\end{eqnarray}
\begin{eqnarray}	
&&  V^{(n)}\;\simeq\;\frac{N}{Z}\,V^{(p)}\,,
\quad g_{LV(u)}\;=\;1-\frac{8}{3}s_{w}^{2}\quad\text{and}
\quad g_{LV(d)}\;=\;-1+\frac{4}{3}\,s_{w}^{2}\,.
\label{CRtypeIIIc}
\end{eqnarray}
%
An upper bound on
$|\eta_{\mu e}|$  can be derived from the present
experimental upper limit on
the $\mu-e$ conversion rate in the nucleus of $^{48}_{22}$Ti,
 ${\rm CR}(\mu\,{\rm Ti}\to e\, {\rm Ti})\lesssim 4.3\times 10^{-12}$.
From eqs.~(\ref{CRtypeIII})-(\ref{CRtypeIIIc})
we get:
\begin{equation}
|\eta_{\mu e}|\;\lesssim\; 2.6\times 10^{-7}\,.
\end{equation}
%
If in the $\mu-e$ conversion experiments with $^{48}_{22}$Ti
the prospective sensitivity to
${\rm CR}(\mu\,{\rm Ti}\to e\, {\rm Ti})\sim 10^{-18}$
will be reached, these experiments will be able to
probe values of  $|\eta_{\mu e}|$ as small as
$|\eta_{\mu e}|\sim 1.3\times 10^{-10}$.

\section{Discussion and Conclusions}\label{Conclusions}

We have performed a detailed analysis of
charged lepton flavour violating (LFV) processes $-$ $\mu\to e \gamma$, $\mu\to 3e $ and $\mu-e$ conversion
in nuclei $-$ in the context of see-saw type extensions of the Standard Model, in which the scale of new physics $\Lambda$ is
taken in the TeV range, $\Lambda\sim (100-1000)$ GeV. In this class of models, an effective Majorana mass term
for the light left-handed active neutrinos is generated after electroweak symmetry breaking due to the decoupling of additional
``heavy'' scalar and/or fermion representations. We have analyzed in full generality the phenomenology
of the three different and well-known (see-saw) mechanisms of neutrino mass generation, in their
minimal formulation: $i)$ type I see-saw models, in which the new particle content consist of  at least 2 RH neutrinos,
which are not charged under the SM gauge group;  $ii)$ type III see-saw models, where the RH neutrinos
are taken in the adjoint representation of $SU(2)_{L}$ with zero hypercharge;
$iii)$ type II see-saw (or Higgs triplet) models, where the scalar sector of the theory is extended with the addition of at least one
scalar triplet of $SU(2)_{L}$  coupled to charged leptons.

Under certain conditions the couplings of the SM charged leptons with the  new  fermions
and/or scalars are, in principle, sizable enough to allow for their production and detection at present collider facilities, LHC included.
On the other hand, remarkable indirect tests of such scenarios are also possible in ongoing and future experiments looking for
charged lepton flavour violation. Indeed, the flavour structure of the interactions between the SM leptons and the new ``heavy'' particle states
is mainly determined by the requirement of reproducing neutrino oscillation data, in such a way that the unknown parameter
space can be expressed in terms of very few quantities. The latter can, therefore, be
constrained by the measurement of different LFV observables. Further and complementary information is provided
also by experiments which search for lepton number violating phenomena, such as neutrinoless double beta decays of
even-even nuclei.

We summarize below the phenomenological implications  of a possible observation
of the LFV processes given above for each kind of (TeV scale) see-saw extensions of the SM.
\paragraph{ Type I  see-saw results.} In this case
the $\mu\rightarrow e \gamma$ and $\mu\rightarrow 3e$
decay branching ratios ${\rm BR}(\mu\rightarrow e \gamma)$ and
${\rm BR}(\mu\rightarrow 3e)$, and the $\mu - e$ conversion rate in
a  nucleus $\mathcal{N}$,
${\rm CR}(\mu\, \mathcal{N}\to e\, \mathcal{N})$,
$\mathcal{N} = $ Al, Ti, Au,
can have values close to the existing upper limits and
within the sensitivity of the ongoing MEG experiment
searching for the $\mu\rightarrow e \gamma$ decay and
the future planned $\mu - e$ conversion and
$\mu\rightarrow 3e$ decay  experiments
\cite{comet,mu2e,PRIME,projectX,muto3eNext}.
The relevant LFV observable in the minimal scenario, with the
addition of only two RH neutrinos to the SM particle content,
is provided by the quantity $|(RV)_{\mu1}^{*} (RV)_{e1}|$,
where $(RV)_{\ell j}$ ($j=1,2$) denote
the couplings of the fermion singlets to the SM charged
leptons (see eqs.~(\ref{mixing-vs-y}) and (\ref{mixing-vs-yIH})).
If the MEG experiment reaches the projected
sensitivity and no positive signal will be observed
implying that ${\rm BR}(\mu\rightarrow e \gamma) < 10^{-13}$,
there still will be a relatively large interval
of values of $|(RV)_{\mu1}^{*} (RV)_{e1}|$,
as Fig. \ref{fig22} shows, for
which the  $\mu - e$ conversion and
$\mu\rightarrow 3e$ decay
are predicted to have observable rates in the planned
next generation of experiments.

 It follows from the analysis performed by us that
as a consequence of an accidental cancellation
between the contributions due to the different one loop
diagrams in the $\mu - e$ conversion amplitude,
the  rate of $\mu - e$ conversion in Al and Ti or in Au
can be strongly suppressed for certain values of the
see-saw scale $M_1$. As we have seen,
this suppression can be efficient either for
the conversion in Al and Ti
or for the conversion in Au, but not for
all the three nuclei, the reason
being that the values of
$M_1$ for which it happens in  Al and Ti
differ significantly from those for which
it occurs in Au. In both the cases of
Al or Ti and Au, the suppression
can be effective only for values of $M_1$ lying
in very narrow intervals
(see  Figs. \ref{Cmue1} and \ref{CRvsBRs}).

In the case of IH light neutrino mass spectrum,
all the three LFV observables, ${\rm BR}(\mu\rightarrow e \gamma)$,
${\rm BR}(\mu\rightarrow 3e)$ and
${\rm CR}(\mu\, \mathcal{N}\to e\, \mathcal{N})$,
can be strongly suppressed due to the fact that
the LFV factor $|(RV)_{\mu1}|^2 \propto
|U_{\mu 2}+i\sqrt{m_{1}/m_{2}}U_{\mu 1}|^{2}
\cong |U_{\mu 2}+iU_{\mu 1}|^{2}$,
in the expressions of the three rates
can be exceedingly small.
This requires a special relation
between the Dirac
and the Majorana CPV phases  $\delta$ and $\alpha_{21}$,
as well as between the neutrino mixing angle
$\theta_{13}$ and the phase $\delta$ (see eq.~(\ref{s13min2})).
For the values of $\sin\theta_{13}$ from the
current $3\sigma$ allowed interval, eq.~(\ref{DBayth13}),
one can have $|U_{\mu 2}+iU_{\mu 1}|^{2}\cong 0$
provided $0\leq \delta \ltap 0.7$.
{\it A priori} it is not clear why the relations
between $\delta$ and $\alpha_{21}$,
and between $\delta$ and $\theta_{13}$,
leading to  $|U_{\mu 2}+iU_{\mu 1}|^{2} = 0$,
should take place
(although, in general, it might be a
consequence of the existence of an
approximate symmetry). The suppression
under discussion cannot hold if, for instance,
it is experimentally established that
$\delta$ is definitely bigger
than 1.0. That would be the case
if the existing indications
\cite{Fogli:2011qn} that $\cos\delta < 0$
receive unambiguous confirmation.

 We note finally that for  $M_1\gtap 100$ GeV we have: 
${\rm BR}(\mu\rightarrow 3e)/{\rm BR}(\mu\rightarrow e \gamma) \gtap 0.031$.
Thus, if it is experimentally established that
${\rm BR}(\mu\rightarrow 3e)/{\rm BR}(\mu\rightarrow e \gamma)$
is definetely smaller than the quoted lower bound,
the model considered with $M_1\gtap 100$ GeV will be
ruled out. Such a result would be consistent also just
with a see-saw scale $M_1 < 100$ GeV.

\paragraph{Type II see-saw results.}
It follows from the results obtained in  Section~\ref{TypeIISec} that
the predictions for the $\mu \to e \gamma$ and
$\mu \to 3e$ decay branching ratios, as well as
the $\mu - e$ conversion rate in a nucleus $\mathcal{N}$,
in the TeV scale type II see-saw scenario considered
exhibit, in general, different dependence
on  the masses of the singly- and doubly-charged Higgs
particles $\Delta^{+}$ and $\Delta^{++}$,
which mediate (to leading order) the three processes.
For $m_{\Delta^{+}}\cong m_{\Delta^{++}} \cong M_{\Delta}$,
all the three rates are proportional to $M^{-4}_{\Delta}$,
i.e., they diminish as the 4th power of the see-saw scale
when the latter increases.

   The matrix of Yukawa couplings $h_{\ell\ell^\prime}$ which
are responsible for the LFV processes of interest, is directly
related to the neutrino Majorana mass matrix and thus to
the PMNS neutrino mixing matrix $U$.
As a consequence, ${\rm BR}(\mu\rightarrow e \gamma)$,
${\rm BR}(\mu\rightarrow 3e)$ and  ${\rm CR}(\mu\, \mathcal{N}\to e\, \mathcal{N})$
depend, in general, on the neutrino mass and mixing parameters,
including the CPV phases in $U$.

To be more specific,  ${\rm BR}(\mu\rightarrow e \gamma)$
does not depend on the Majorana CPV phases and on ${\rm min}(m_j)$,
and its dependence on the Dirac CPV phase and on the type of neutrino
mass spectrum is insignificant.
In contrast, both ${\rm BR}(\mu\rightarrow 3e)$ and
${\rm CR}(\mu\, \mathcal{N}\to e\, \mathcal{N})$
exhibit very strong dependence on the
type of neutrino mass spectrum and on the values
of the Majorana and Dirac CPV phases.
As a consequence, the predictions for
${\rm BR}(\mu\rightarrow 3e)$ and
${\rm CR}(\mu\, \mathcal{N}\to e\, \mathcal{N})$
for given
$M_{\Delta}$ can vary by
several orders of magnitude not only when
the spectrum changes from NH (IH) to QD
as a function of the lightest neutrino mass,
but also when one varies only the values of the
CPV phases keeping the type of the
neutrino mass spectrum fixed.
All the three observables under discussion
can have values within the sensitivity of the
currently running MEG experiment on
the $\mu\rightarrow e \gamma$ decay and
the planned future experiments
on  the $\mu\rightarrow 3e$  decay and
$\mu - e$ conversion. However,
for a given see-saw scale in the range
of $\sim (100 - 1000)$ GeV, the planned
experiments on $\mu - e$ conversion in Al or Ti
will provide the most sensitive probe of the
LFV Yukawa couplings of the TeV scale type II
see-saw model.

\paragraph{Type III see-saw results.} Unlike the type I see-saw extension
of the SM discussed in Section~\ref{TypeISec}, in this scenario we have
several $-$ possibly sizable $-$ lepton flavour violating interactions in the low energy effective Lagrangian,
due to the higher $SU(2)_{L}$ representation of the new fermion fields.
In particular, FCNCs arise at tree-level from the non-unitarity of the PMNS matrix
(see eq.~(\ref{ellNCIII})). Thus, the effective $\mu-e-Z$ coupling in (\ref{ellNCIII}) makes it possible an enhancement
of at least two orders of magnitude of the rates of $\mu\to e \gamma$, $\mu\to 3e$
and $\mu-e$ conversion with respect to the  ones predicted in the type I see-saw scenario,
with RH neutrinos taken in the TeV range.  Consequently, all the predicted LFV observables may be probed
in the related present and future experiments.  As in the previous scenarios, the strongest constraint
on the flavour structure of this class of models is by far provided by the expected very high sensitivity reach
of $\mu-e$ conversion experiments.
\newline

 In conclusion,  the oncoming combination of data on neutrino oscillations, collider searches and lepton number/flavour violating processes
represent an important opportunity to reveal in the next future the fundamental mechanism at the basis of the generation of neutrino masses
as well as the underlying physics beyond the standard theory.

\section*{Acknowledgments}
This work was supported in part by the INFN program on
``Astroparticle Physics'', by the Italian MIUR program on
``Neutrinos, Dark Matter and  Dark Energy in the Era of LHC''
(D.N.D. and S.T.P.) and by the World Premier International
Research Center Initiative (WPI Initiative), MEXT,
Japan  (S.T.P.),
by the DFG cluster of excellence ``Origin and Structure of the Universe''
and the ERC Advanced Grant project ``FLAVOUR''(267104) (A.I.),
and by the Funda\c{c}\~{a}o para a Ci\^{e}ncia e a
Tecnologia (FCT, Portugal) through the projects
PTDC/FIS/098188/2008,  CERN/FP/116328/2010
and CFTP-FCT Unit 777,
which are partially funded through POCTI (FEDER) (E.M.).

\end{document}